\documentclass[aps,prl,twocolumn,tightenlines,superscriptaddress,amsmath,amssymb]{revtex4-1}
\usepackage{graphicx}
\usepackage{bm}

\usepackage{multirow}

\usepackage{grffile}

 \newcommand{\be}{\begin{equation}}
 \newcommand{\ee}{\end{equation}}
\newcommand{\bea}{\begin{eqnarray}}
\newcommand{\eea}{\end{eqnarray}}
\newcommand{\ba}{\begin{eqnarray*}}
\newcommand{\ea}{\end{eqnarray*}}

\newcommand{\bx}{\mathbf{x}}

\newcommand{\Tr}{\mathrm{Tr}}

\newcommand{\m}[1]{\mathcal{#1}}

\usepackage{multirow}

\usepackage{color}
\usepackage[english]{babel}
\usepackage{dcolumn}

\begin{document}

\title{Understanding Quantum  Tunneling through Quantum Monte Carlo Simulations}

\author{Sergei V. Isakov}
 \affiliation{Google, 8002 Zurich, Switzerland}
\author{Guglielmo Mazzola}
 \affiliation{Theoretische Physik, ETH Zurich, 8093 Zurich, Switzerland}
 \author{Vadim N. Smelyanskiy}
 \affiliation{Google, Venice, CA 90291, USA}
 \author{Zhang Jiang}
 \affiliation{NASA Ames Research Center, Moffett Field, CA 94035, USA}
\author{Sergio Boixo}
 \affiliation{Google, Venice, CA 90291, USA}
\author{Hartmut Neven}
 \affiliation{Google, Venice, CA 90291, USA}
\author{Matthias Troyer}
 \affiliation{Theoretische Physik, ETH Zurich, 8093 Zurich, Switzerland}

             
\begin{abstract}
The tunneling between the two ground states of an Ising ferromagnet is a typical example of many-body tunneling processes between two local minima, as they occur during quantum annealing. 
Performing quantum Monte Carlo (QMC) simulations we find that the QMC tunneling rate displays the same scaling with  system size, as the rate of incoherent tunneling. The scaling in both cases is $O(\Delta^2)$, where $\Delta$ is the tunneling splitting. An important consequence is that QMC simulations can be used to  predict the performance of a quantum annealer for tunneling through a barrier. Furthermore, by using open instead of periodic boundary conditions in imaginary time, equivalent to a projector QMC algorithm, we obtain a quadratic speedup for QMC, and achieve linear scaling in $\Delta$. 
We provide a physical understanding of these results and their range of applicability based on an instanton picture.  
\end{abstract}
  \maketitle

  Quantum annealing~\cite{PhysRevE.58.5355,Farhi20042001,Santoro29032002,boixo2014,Ronnow:2014fd,King:2015vl} (QA) has been proposed as a method to solve combinatorial optimization problems. In contrast to its closely related classical counterpart, simulated annealing (SA)~\cite{Kirkpatrick13051983}, which makes use of thermal fluctuations to escape  local minima of the energy landscape in the search for a low energy solution, QA can additionally exploit quantum tunneling. 
In QA the system closely follows the ground state of a time-dependent Hamiltonian $H(t)$ whose  initial ground state at $t=0$  is easy to prepare. The final Hamiltonian $H(t_{\rm final})$ encodes the cost function of a combinatorial optimization problem.  

Random ensembles of hard problems are closely connected to spin glass models known in statistical physics. 
There one typically passes through a second order quantum phase transition from a paramagnetic into a glassy phase, where the energy gap closes polynomially with problem size $N$, and then encounters a cascade of avoided level crossings with typically exponentially small gaps $\Delta \propto e^{-\alpha N}$ inside the glassy phase \cite{Santoro29032002,Altshuler13072010,Farhi2012,knysh2015}. 
 In other problems, such as Grover search or number partitioning, there may  just be a single exponentially small gap at a single first order quantum phase transition. 
Avoided level crossings with exponential gaps are the main bottleneck in quantum annealing and are in most cases associated with tunneling processes.


 Simulations  are  important to  understand the mechanisms of QA and find the class of problems for which QA may perform better than SA and other classical algorithms. QMC simulations have been performed \cite{Santoro29032002,PhysRevB.66.094203,Heim12032015} on problems of much larger sizes than accessible by direct integration of the time dependent Schr\"odinger equation. In particular, a recent numerical study of random Ising spin glass instances \cite{Heim12032015} has  reconciled  expectations of quantum speedup  based on QMC simulations \cite{Santoro29032002} with experiments that failed to detect it \cite{Ronnow:2014fd}.

The major bottlenecks of QMC simulations of QA are also associated with tunneling events. However, while QMC faithfully samples the equilibrium thermal state of a quantum system it does not  directly simulate its unitary time evolution. In particular, the universal critical exponents at second order quantum phase transitions are different than those of the stochastic QMC dynamics, as was recently pointed out in this context \cite{PhysRevLett.114.147203}. 
Nevertheless,  correlations between QMC dynamics and the gap have recently been  observed in simple models \cite{brady2015quantum}.

In this Letter we show that despite the different dynamics there exists a broad class of tunneling problems where QMC is not ``merely''  a quantum-inspired classical optimization algorithm~\cite{Santoro29032002,boixo2014,Ronnow:2014fd}. In these problems the time of QMC to simulate quantum mechanical tunneling scales identically (in leading exponential order)  with the problem size to the tunneling rate of a physical system and QMC is thus a quantitatively faithful predictor  of QA performance. We also discuss possible types of problems where this may not apply.

\emph{Tunneling decay of a metastable state} --- To gain insight into the equivalence of QA and QMC we consider the tunneling between two nearly degenerate minima $\bx_1$ and $\bx_2$ of a potential $V({\bf x})$. The pioneering work of Langer~\cite{langer_theory_1967,langer_statistical_1969} makes an explicit connection between the tunneling rate of a particle and the classical
Kramers escape rate from the metastable state of a non-linear stochastic field process. This  sheds light on how QA and QMC tunneling dynamics are related.

Within a semiclassical picture, the wave function decays exponentially in the classically forbidden region. In the particular case where the action under the barrier is purely imaginary this corresponds to a particle moving
with imaginary momentum
along the imaginary time axis $t = -i \tau$~\cite{coleman_fate_1977,takada_wentzelkramersbrillouin_1994}. The  amplitude $\Delta$ of tunneling from the ground state associated with a local minimum $V(\bx_1)=0$ is determined by the path integral $  K_\tau(\bx', \bx_1)$~=~$\int D[\bx(\tau')]\; e^{- {\m S_\tau[\bx(\tau')]}/{\hbar}}$
between the local minimum $\bx(0) \approx \bx_1$ and the turning point $\bx(\tau) = \bx'$ at the barrier exit chosen to maximize $K_\tau$. Here $\m S_\tau[\bx(\tau')] =  \int_{0}^{\tau} d\tau'\frac 1 2 m \dot \bx ^2(\tau') + V(\bx(\tau'))$ is the action of the path under the barrier and $\tau \to \infty$. 
The dominant contribution comes from the stationary action path (instanton) $\bx^*(\tau')$ corresponding to a particle moving in the inverted potential $-V(\bx)$. The tunneling amplitude is
  $\Delta \propto \exp({-\m S_\tau[\bx^*(\tau')]/\hbar})$.

Similar arguments are known in statistical physics where the partition function $\m Z = \Tr \,K_\beta$ of a state thermalized near a local minimum $\bx_1$ of the potential corresponds to the path integral in imaginary time with periodic boundary conditions for $\bx(\tau)$  with $0 \le \tau < \beta = \hbar/k_B T$. By  tunneling away from the minimum, the partition function acquires an imaginary part. It is dominated by the instanton/anti-instanton path  
$\bx^{**}(\tau)$ 
that moves under the barrier starting near $\bx_1$, reaches the turning point $\bx'$, and returns ~\cite{0038-5670-25-4-R01}. We note that $-2\,{\rm Im} \m Z/(\beta\, {\rm Re}\m Z)  \propto \Delta^2$~\cite{Chudnovsky-book}  gives a {\it squared} tunneling amplitude ($\propto e^{-\m S_\beta[x^{**}(\tau)]/\hbar}$) because we pay the cost of creating an instanton and an anti-instanton.

In the context of QA we introduce a slowly varying field that changes the order of the minima $\bx_{1,2}$ of $V(\bx, t)$. At the start of QA the system is localized in the vicinity of $\bx_1$ and at the end it arrives at the vicinity of $\bx_2$ after a tunneling event at time $t_c$ when $V(\bx_1,t_c) \approx V(\bx_2,t_c)$. 
In the case of open system QA when the dephasing time $T_2\ll \Delta^{-1}$ there is incoherent tunneling at $t\approx t_c$  from $\bx_1$ to $\bx_2$ with rate $ T_2\,\Delta^2$, determining the time scale of open system QA~\cite{amin_macroscopic_2008,Amin-Averin-Nesteroff-2009,boixo_computational_2014}. The same scaling with $\Delta^2$ is also obtained in closed systems by the Landau-Zener formula.

Following Refs.~\cite{langer_theory_1967,langer_statistical_1969}, the tunneling decay rate {\it formally} corresponds to the Kramers escape rate from a metastable state of a classical 1D field with order parameter $\bx(\tau)$ satisfying the periodic boundary conditions and free energy functional $\m F = \m S_\beta[\bx(\tau)]/\beta $. 
The stochastic evolution of this field $\bx(\tau, t)$ in time $t$ is described by the Langevin equation $ \bx_t =-\mu \beta^{-1}[ {\bm \nabla} V -m  \bx_{\tau\tau}] + (2 \mu \hbar/\beta)^{1/2}\, {\bm \eta}$, where ${\bm \eta}(\tau,t)$ is a random force delta correlated in both of its arguments and $\mu$ is a relaxation coefficient. We now observe that the same dynamics describes the standard path integral QMC to calculate the partition function $\m Z$. QMC samples paths $\bx(\tau,t)$ with sweeping rate $\propto \mu$. The functional $\m F[\bx(\tau)]$ has a saddle point $\bx^{**}(\tau)$ that the QMC trajectory $\bx(\tau,t)$ crosses during the escape event from the metastable state $\bx_1$ towards $\bx_2$. According to Kramers theory the escape rate is $W \propto e^{- \m F / k_B T} = e^{-\m S_\beta[\bx^{**}(\tau)]/\hbar}$. This saddle point is precisely the instanton/anti-instanton path, and therefore the QMC escape rate  $W \propto \Delta^2 \propto -{\rm Im}\, \m Z $. Therefore, in this archetypical example, the time needed for a physical system to tunnel is equal, within exponential accuracy, to the corresponding simulation time of quantum Monte Carlo. 

We validated these arguments by simulations of tunneling in a one-dimensional double well potential $V(x)=\lambda\, x^4 - x^2$, where  $\Delta$  depends exponentially on $\lambda$.
Performing QMC simulations in continuous space \cite{RevModPhys.67.279} we compared the average QMC tunneling time to $1/\Delta^2$ and find excellent agreeement over a wide range of time scales (see Supplementary Materials (SM)). Furthermore, we find that the QMC scaling does not significantly depend  on whether local or global updates are used. Using  open instead of periodic boundary conditions in imaginary time describes a so-called path integral ground state (PIGS) simulation~\cite{pigs}. There, the tunneling trajectory is dominated by configurations with a single instanton $\bx^*(\tau)$ with corresponding escape rate $W \propto \Delta$ -- giving quadratic speedup of QMC over incoherent tunneling.



\emph{Tunneling in a transverse field Ising model} --- We now show that our findings are not limited to continuous variables, but extend to tunneling through barriers in quantum spin systems. 
To establish the equivalence of QA and QMC tunneling dynamics in this case we study the archetypical model of an Ising ferromagnet in the presence of a weak transverse field $\Gamma$ with Hamiltonian $H=-\Gamma \sum \sigma_j^x - \sum J_{ij} \sigma_i^z \sigma_j^z$, considering both a linear chain with couplings $J_{ij}=\delta_{i,j+1}+\delta_{i,j-1}$ and fully connected clusters with $J_{ij}=1/2L$, where $L$ is the number of spins. 

For small $\Gamma$ there are two degenerate ground states:  the configuration labeled $|\!\uparrow\rangle$ with spins aligned pointing (predominantly) up and average magnetization per site  $m \equiv \frac{1}{L}\sum_i\langle \sigma_i^z\rangle \approx +1$  and the configuration labeled $| \!\downarrow \rangle$  with  $m \approx -1$.
For finite $L$ the transverse field term mixes the two states with an exponentially small (in $L$)  but nonzero  tunneling matrix element $\epsilon=  \langle \uparrow\! | H | \!\downarrow\rangle$. This lifts the degeneracy between the two ground states, resulting in an exponentially small energy gap $\Delta=2\epsilon$ between the  states $| \psi_\pm \rangle = 1 / \sqrt{2} \left( |\!\uparrow \rangle \pm | \!\downarrow \rangle \right)$. 

  \begin{figure}[tb]
 \includegraphics[width=1.0\columnwidth]{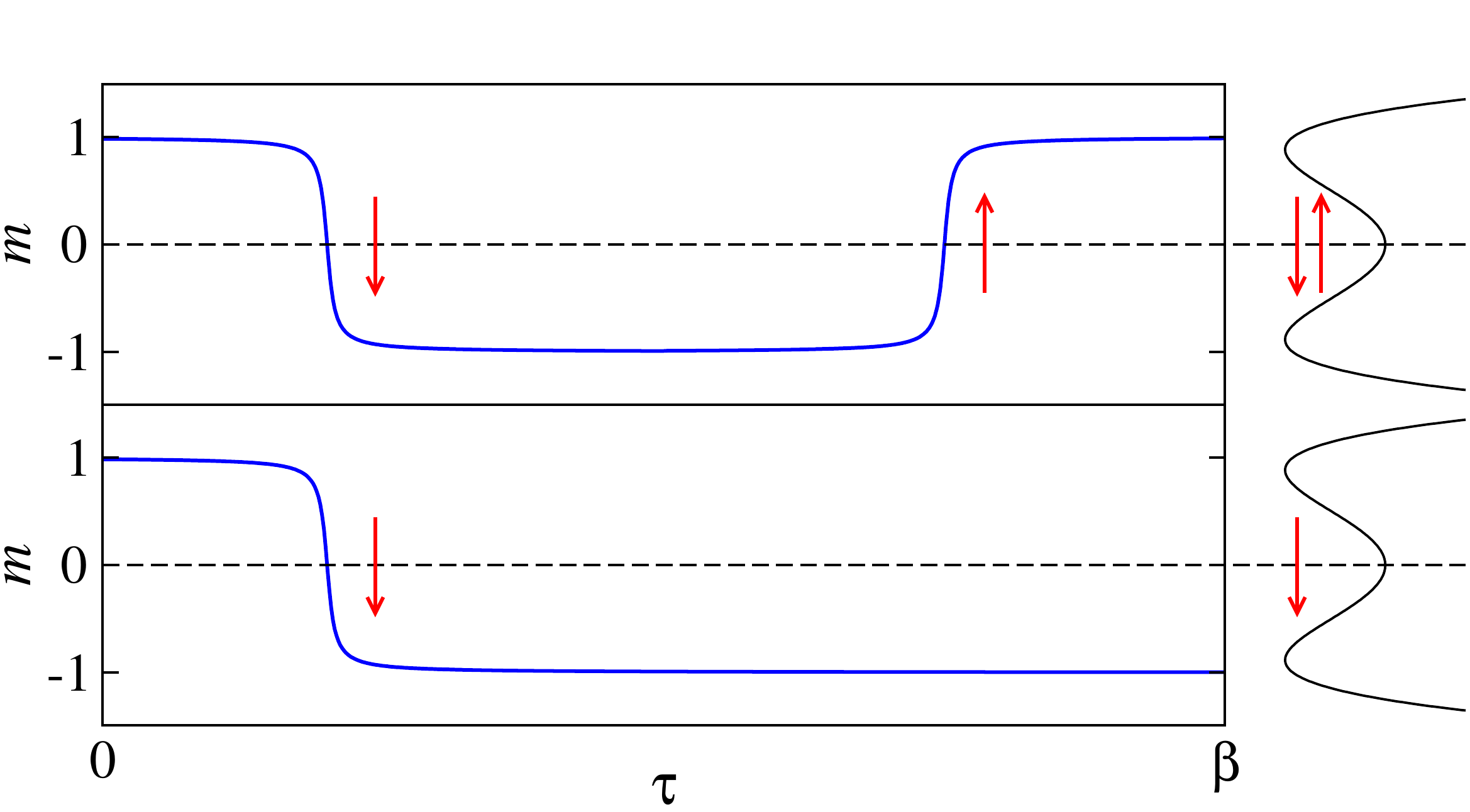}
 \caption{ \emph{Upper panel:} Typical instanton/anti-instanton trajectory with periodic boundary condition in imaginary time. Along such trajectories the path samples both the $| \!\uparrow~ \rangle$ and $| \!\downarrow~ \rangle$ states. The magnetization $m$ exhibits two {jumps} corresponding to instantons (tunneling events) in imaginary time (red arrows). \emph{Lower panel:} with open boundary condition only one instanton is required and the tunneling probability thus increased. On the right we sketch the double-well potential. 
 }
 \label{fig:instanton}
 \end{figure}

Adiabatically tuning a (weak) longitudinal field $-h\sum_i\sigma_i^z$ from a small positive to a small negative value, we encounter a tunneling problem, which is similar to the typical tunneling problem of QA at avoided level crossings, with the spins having to tunnel from  $|\!\uparrow \rangle$ to $|\!\downarrow \rangle$.
This event is described by an instanton path $m = m(\tau)$ in which all spins evolve in an identical fashion (see SM). The instanton dynamics can be described by an effective double well model, whose degenerate minima are located at $m=\pm1$ (see Fig.~\ref{fig:instanton}). 

\begin{figure}[b]
\includegraphics[width=1.0\columnwidth]{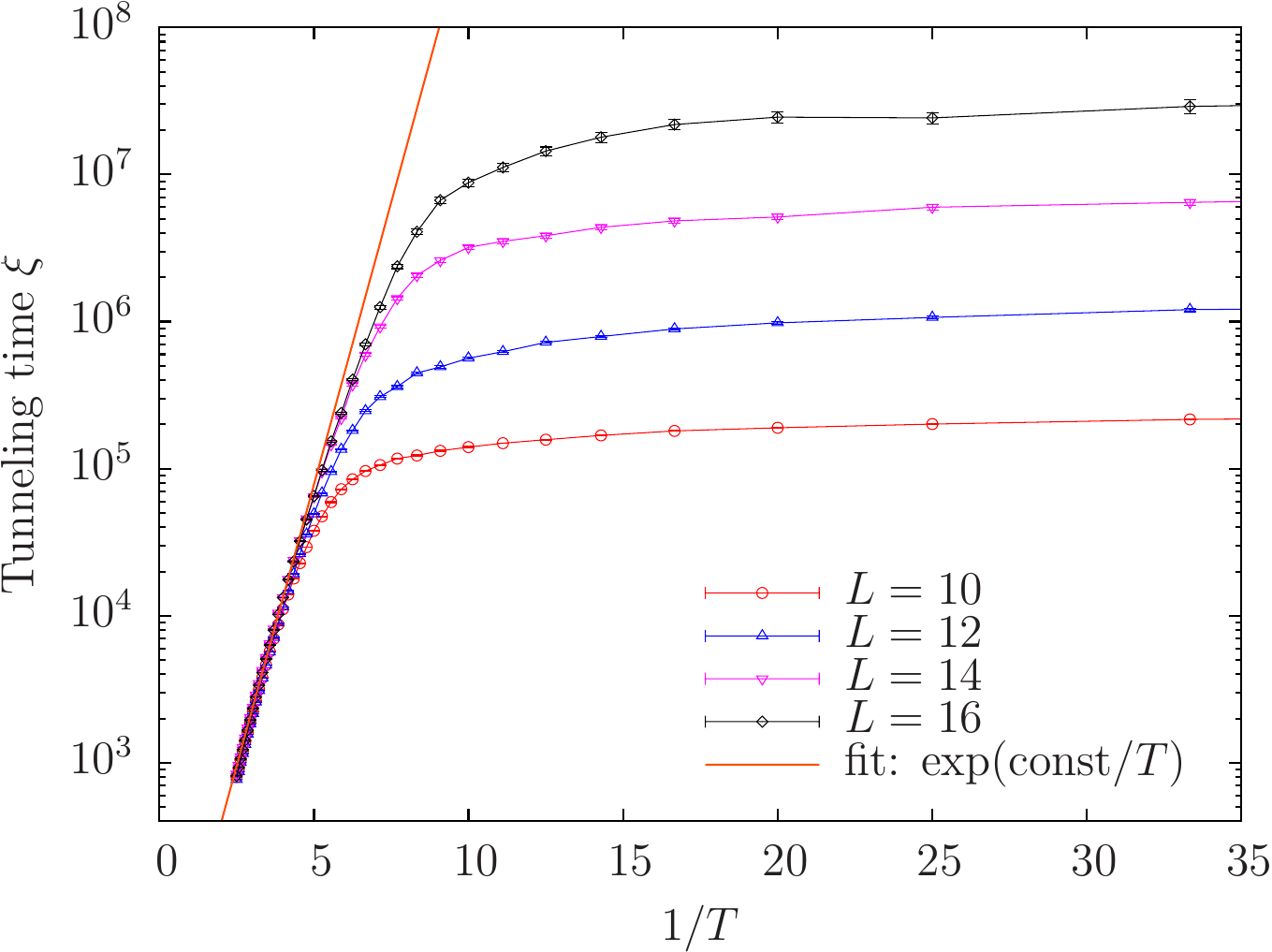}
\caption {
Path integral QMC transition time $\xi$ as a function of the inverse temperature for a chain at $\Gamma=0.7$. At high temperatures $\xi$  is independent of the system size $L$ and suppressed as $\exp(-E_{\rm barrier}/T)$. At low temperatures, quantum tunneling becomes more efficient and depends only on $L$.
}
\label{fig:tunneling:t}
\end{figure}

We note that coupling to an environment could lead to thermally activated events, whose pathways traverses the high energy states $m\approx0$ with energy $E_{\rm barrier}  \sim L/2$ for a fully connected cluster and $E_{\rm barrier}=4$ for a linear chain. These are suppressed by a factor $\exp(-\beta E_{\rm barrier})$ and at low temperatures quantum tunneling becomes advantageous, as shown in Fig.~\ref{fig:tunneling:t}).

\emph{Tunneling in path integral QMC} --- QMC simulations are performed by sampling  imaginary time paths, obtained from a mapping of the partition function of a $D$-dimensional quantum system to a $(D+1)$-dimensional classical path integral configuration. For the transverse field Ising model this is just a classical Ising model in $D+1$ dimensions.
-
The path integral configurations consist of $P$ replicas of the same physical system with spins $S_{i,\tau}= \pm 1$  where the index  $i=1,\cdots,L$ denotes the spatial index and the index  $\tau=1,\cdots,P$ labels the time slices. 
Our QMC simulations have been performed both with a large number of replicas $P=128$ in order to be close to the physical continuous time limit, and also directly in the continuous time limit \cite{Eur.Phys.J.B.9.233}. For updates we use variants of the Wolff \cite{PhysRevLett.62.361} and Swendsen-Wang algorithm \cite{PhysRevLett.58.86} to build local (in space) clusters along the imaginary time direction \cite{PhysRevB.68.104409}.

During the QMC simulation the total magnetization $m(\tau)\equiv \frac{1}{L}\sum_{i=1}^L S_{i,\tau}$ evolves stochastically in Monte Carlo time $t$. Preparing the  system in the vicinity of a local minimum, for example by choosing $S_{i,\tau}=1$, most of the time all  replicas sample spin configurations which are fluctuations around  the same minimum energy configuration, i.e. $ m(\tau)\approx1$. Every now and then the path $m(\tau,t)$ evolves towards a transition state $m^{**}(\tau)$ corresponding to the saddle point of the free energy functional $\m F[m(\tau)]$ of the classical spin model. Similar to the discussion in the continuous case, this saddle point corresponds to an instanton/anti-instanton pair (see Fig.~\ref{fig:instanton} and SM). The instantonic path $m^{**}(\tau)$ 
costs energy as it creates two domain walls in the imaginary time axis, which separate replicas which opposite magnetization $m(\tau)$. These domain walls can diffuse  in opposite directions around the imaginary time loop, changing the total manetization to $m(\tau)=-1$ $\forall\, \tau$ when they annihilate, signaling the completion of a tunneling event. The creation of $m^{**}(\tau)$ represents the rate-limiting process of tunneling decay in both QMC and QA  whose rate is $\propto\Delta^2$.

To measure the tunneling time $\xi$ we start QMC simulations in a fully polarized state with $m(\tau)=1$ and measure the number of QMC sweeps (defined as one attempted update per spin) required to obtain a well separated instanton/anti-instanton pair. We detect the latter by requiring that at least the $25\%$ of the replicas reverse their magnetization to $m(\tau) = -1$~\footnote{The number 25\% is somewhat arbitrary but we find that waiting until all, $25\%$ or $2\%$ of the replicas reverse magnetization makes only small differences.}.

  \begin{figure}[t]
 \includegraphics[width=1.0\columnwidth]{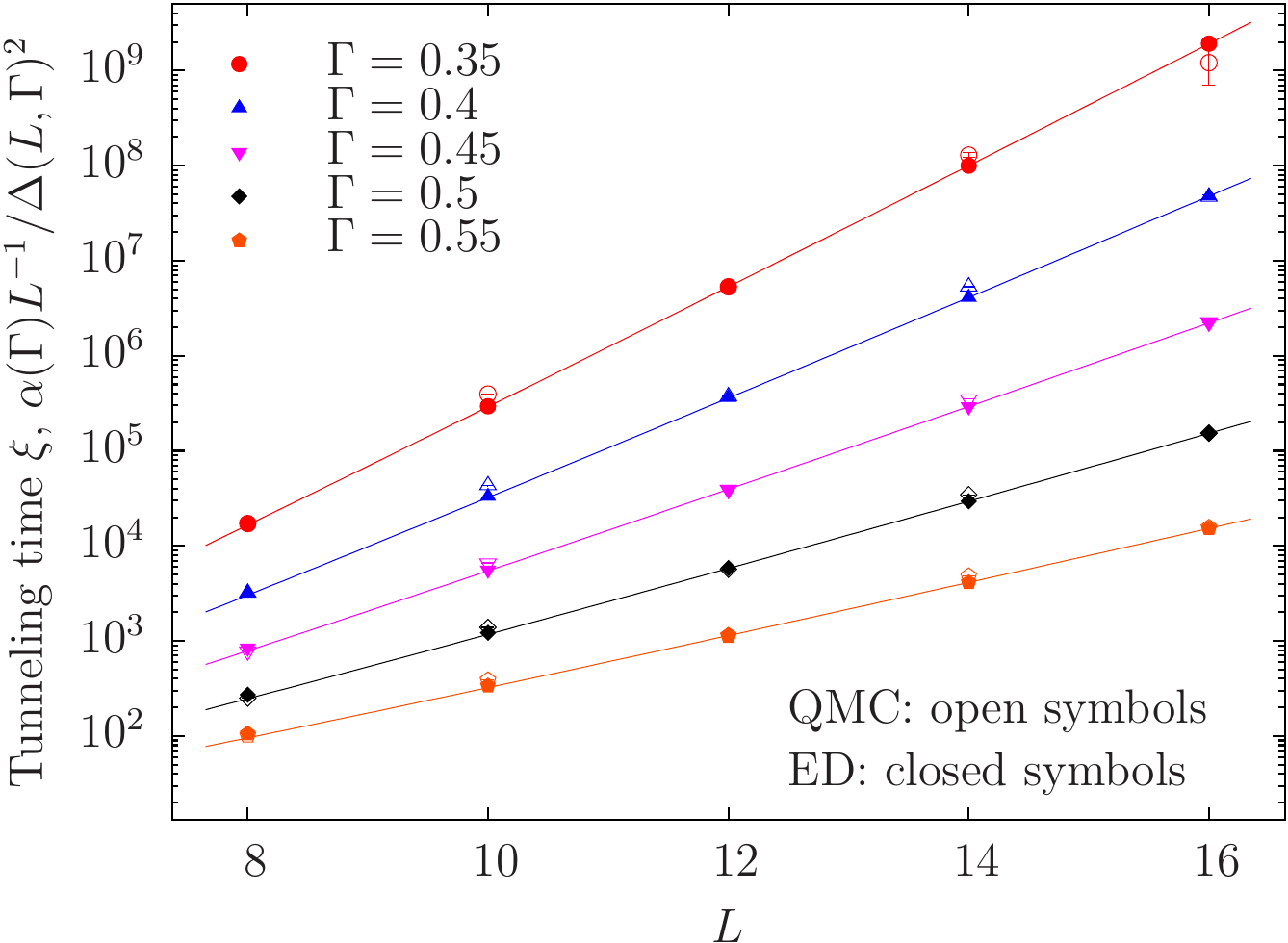}
 
\caption{
Average QMC tunneling time $\xi$ as a function of system size $L$ (open symbols) for a fully connected graph at $\beta=8$ for various values of the transverse field $\Gamma$. Exponential fits of the times for $12 \le L \le 16 $ are shown as solid lines. To compare to physical QA we also show $\alpha(\Gamma)/\Delta(\Gamma,L)^2$, obtained by exact diagonalization (ED). Rescaling by $L$-independent constants $\alpha(\Gamma)$ we find identical scaling with system size $L$.
 }
 \label{fig:pbc}
 \end{figure}

%

In Fig.~\ref{fig:pbc} we show the measured  average tunnelling time $\xi$ in QMC fully connected clusters as a function of $L$ and $\Gamma$ and observe an exponential scaling with $L$.  There is only a very weak temperature dependence for QMC in the low temperature quantum regime, mostly due to
instanton diffusion.
As shown in Fig.~\ref{fig:pbc} the scaling of QMC simulations is identical to $1/\Delta^2$ within error bars, thus confirming the identical scaling behavior of both types of dynamics. The same behavior is observed for linear chains (see SM).



  \begin{figure}[t]
 \includegraphics[width=1.0\columnwidth]{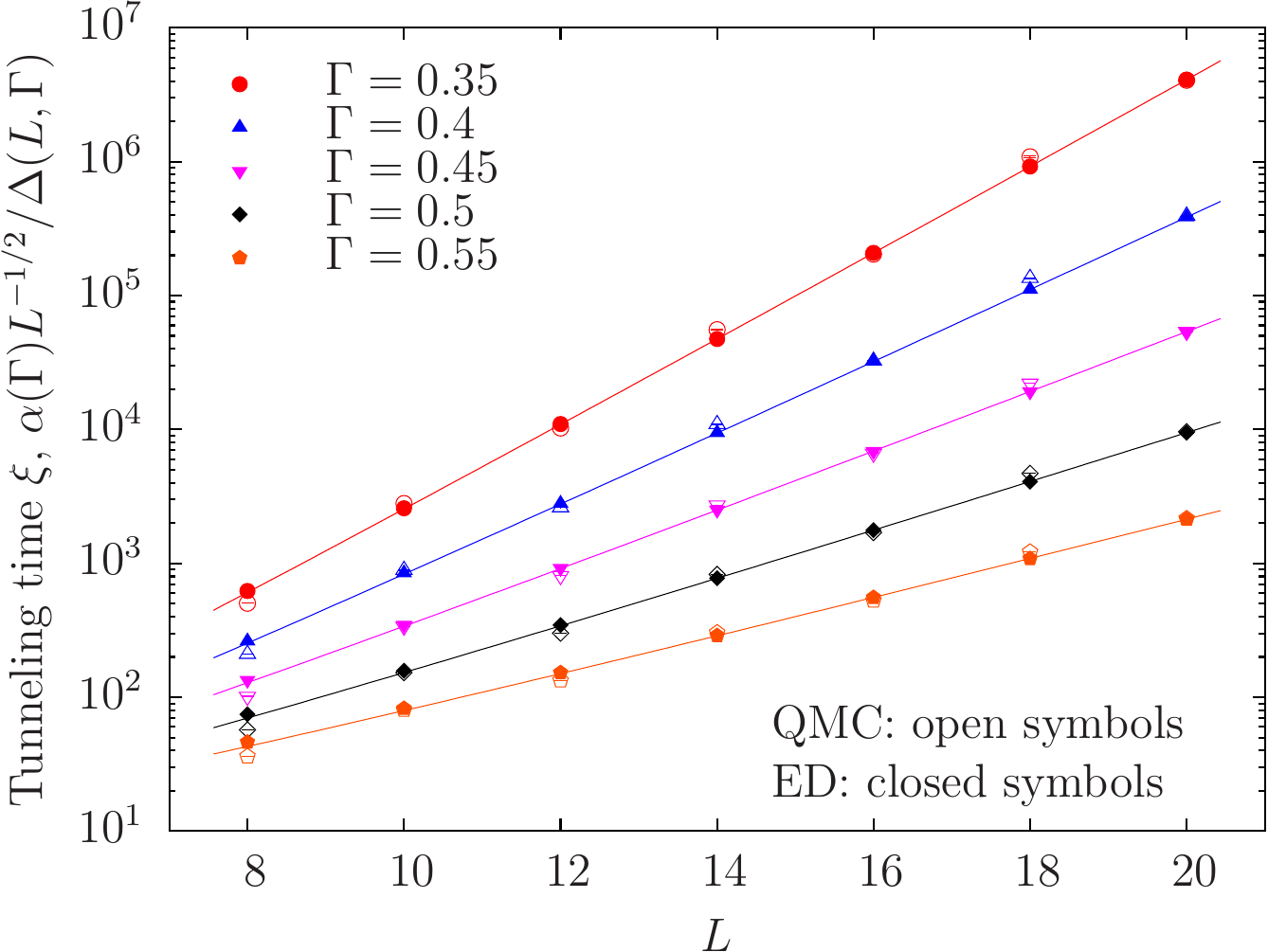}
 \caption{ (color online). 
 Average PIGS tunneling  time $\xi$ as a function of  $L$ (open symbols)  for fully connected graph at $\beta=8$ with various values of the transverse field $\Gamma$. Exponential fits of the times for $12 \le L \le 18$ are shown as solid lines. Points proportional to $1/\Delta (L)$, obtained with exact diagonalization, are shown with filled symbols. }
 \label{fig:obc}
 \end{figure}

\emph{Accelerating tunneling in QMC} --- Similar to the continuous case, we expect a quadratic speedup for PIGS simulations with open boundary conditions as in this case only one domain wall (instanton) is created in the magnetization reversing process (see Fig.~\ref{fig:instanton}). Indeed, as shown in  Fig.~\ref{fig:obc} and Table I of SM the scaling exponent is reduced by a factor of two compared to QMC with periodic boundary conditions and open system QA, and the tunneling time now scales as $1/\Delta$ instead of $1/\Delta^2$.  PIGS algorithm can be viewed as projecting from a trial state (given by the boundary conditions in imaginary time) and  sampling from the ground state wave function at large enough $\beta$, hence providing the tunneling probability amplitude $\Delta$. 
This finding may explain the recently observed superiority of QMC projecting techniques compared to PIMC in continous space models \cite{inack2015simulated}.

\emph{Potential obstructions for QMC} --- While our findings apply to tunneling in a broad class of mean field models with purely imaginery time instantons, we shall also mention several areas where obstructions for the efficient simulation of quantum tunneling with QMC might exist.

QMC sampling may sometimes be inefficient due to topological obstructions such as winding numbers of world lines \cite{hastings}. While PIGS simulations often solve this problem by cutting the periodic boundary conditions, an obstruction remains if the ground state wavefunction and its square are concentrated on different supports~\cite{hastings} -- although suitable trial wave functions at the boundaries of the path integral can alleviate this problem. 
Conversely, the quantum  system might have conserved quantum numbers that limit  tunneling paths to a lower-dimensional subspace than that explored by QMC (see SM).

%

 QMC may also be less efficient in optimization problems that require tunneling to or from multidimensional minima. In such problems 
  the semiclassical action under the barrier $S({\bf x})$ is often not purely imaginary and displays complex features due to the presence of caustics, non-integrability and non-analyticity 
  \cite{huang_wentzel-kramers-brillouin_1990,takada_wentzelkramersbrillouin_1994}. Due to a highly oscillating nature of the wave function 
  in  the classically forbidden region  it is not clear if the tunneling dynamics can be faithfully recovered with QMC. 

Another important open question arises  in  problems  that exhibit a many-body location/delocalization transition  at finite  values of transverse field \cite{Laumann-EPL-2014}. There a delocalized phase can exist in a range of energies  with exponentially many local minima separated by  large Hamming distances and connected by a large number of tunneling paths
 \cite{Laumann-Scardicchio-2014}. QA, in contrast to QMC dynamics, may profit from the positive interference between exponentially many paths.

\emph{Conclusions} --- We conclude by  discussing the consequences of our results for quantum annealing. Despite QMC dynamics being different from unitary evolution, the relevance of instanton configurations for tunneling processes in a class of models with purely imaginery time instantons leads to the same exponential scaling of tunneling rate through a tall barrier in both cases. 
A consequence of this equivalence is that  QMC simulations are predictive of the performance of QA for hard optimization problems where the performance is limited by such tunneling events. 

We also observed that is some cases a version  of QMC with open boundary conditions (PIGS) can provide a quadratic speedup.  While one can, theoretically, recover such a quadratic speedup~\cite{GSadiabatic} in QA  if the evolution of the energy gap is exactly known and if the tunneling is  fully coherent~\cite{Amin-Averin-Nesteroff-2009},  this protocol can hardly be realized in practical  QA.
Nevertheless, a quadratic speed up can be achieved  with polynomial overhead on a universal quantum computer using an  approach  that relaxes the above conditions~\cite{boixo_fast_2010}.

These findings demonstrate that  QMC  simulations can be used as  a powerful and predictive tool to investigate  optimization  problems amenable to quantum annealing. 
Our study demonstrates that  the  quest for quantum speedup using  a physical quantum annealer must  focus on the problems and hardware that allows to reach {\it beyond}  the class of problems discussed in this paper where the identical scaling of QMC and QA  preclude a scaling advantage and where PIGS can achieve a quadratic speedup for tunneling through individual barriers. The absence of calibration and programming errors, the flexibility in simulating arbitrary graph topologies without the need to embed into a hardware graph, and the observed quadratic speedup for tunneling through individual barriers in PIGS simulations makes QMC a competitive classical technology. Nevertheless, a physical QA can still be many orders of magnitude faster than QMC simulations \cite{denchev}.
We expect that the physical mechanisms behind the possible obstructions for QMC discussed above will provide interesting starting points for future studies of potential quantum speedup. In particular, one has to explore the power of QA with non-stoquastic Hamiltonians for which the negative sign problem prevents a matching QMC algorithm.


The work of GM and MT has been supported by the Swiss National Science Foundation through the National Competence Center in Research QSIT and by 
ODNI, 
IARPA
via MIT Lincoln Laboratory Air Force Contract No. FA8721-05-C-0002. 
MT acknowledges  hospitality of the Aspen Center for Physics, supported by NSF grant \#PHY--1066293.
We acknowledge useful discussions with  F. Becca, M. Dykman, and G. Santoro. 

\bibliographystyle{apsrev4-1}

\bibliography{bibQA}

\end{document}


\title[Quantum Monte Carlo vs unitary time dynamics]
{Supplementary Materials for "Understanding Quantum  Tunneling through Quantum Monte Carlo Simulations"}

\author{Sergei V. Isakov}
 \affiliation{Google, 8002 Zurich, Switzerland}
\author{Guglielmo Mazzola}
 \affiliation{Theoretische Physik, ETH Zurich, 8093 Zurich, Switzerland}
 \author{Vadim N. Smelyanskiy}
 \affiliation{Google, Venice, CA 90291, USA}
 \author{Zhang Jiang}
 \affiliation{NASA Ames Research Center, Moffett Field, CA 94035, USA}
\author{Sergio Boixo}
 \affiliation{Google, Venice, CA 90291, USA}
\author{Hartmut Neven}
 \affiliation{Google, Venice, CA 90291, USA}
\author{Matthias Troyer}
 \affiliation{Theoretische Physik, ETH Zurich, 8093 Zurich, Switzerland}

             
\maketitle

\section{Path Integral Monte Carlo in continuous space}

Imaginary time path integral Quantum Monte Carlo  is a very well established technique  for studying quantum statistical mechanics of many-particle realistic systems \cite{RevModPhys.67.279}.
In this section we will briefly introduce path integral QMC in continuous space models.
The formalism is very similar to the one introduced for spin models, the differences coming only from the different types of Hamiltonians.
The typical real space Hamiltonian (in one dimension for sake of simplicity) is given by
\begin{equation}
 H  = {p^2 \over 2 m} + V(x) 
\end{equation}
where $x$ and $p$ are  the position and momentum coordinate respectively, $m$ is the mass of the particle and $V(x)$ is the potential.
Here, the kinetic operator $p^2 / 2 m$  plays the  role of the quantum fluctuation operator as in the transverse field Ising model. The strength of the quantum fluctuations is given by the particle's mass $m$.
By applying the standard Trotter breakup for the density operator $e^{- H/T}$, the partition function at  temperature  $T$ is given by
\begin{align}
\label{eqZ}
 Z = tr~ e^{-H/T} &\approx\int dx  ~\exp [   - \sum_{k=1}^P ( {mPT \over 2}  (x_k - x_{k+1})^2 \notag\\
  &+ {1\over PT}  V(x_k) )]
\end{align}

where $x_k$ is the coordinate of the $k-$th system's replica (time slice) and $P$ is the total number of Trotter replicas. Periodic boundary conditions applies, therefore $x_{P+1} = x_1$.
We refer the reader to Ref.~\cite{RevModPhys.67.279} for additional details concerning energy estimators and more sophisticated Trotter breakups, i.e. more accurate approximate propagators.
Eq.~\ref{eqZ} represents the partition function of a classical ring polymer, made of $P$ beads at the fictitious temperature $PT$. Each replica is connected by harmonic \emph{springs} whose spring constant is given by $\kappa =m (PT)^2$. This harmonic term represents the continuous model analogue of the ferromagnetic  spin interaction between neighboring time slices in the transverse field model (see e.g. Ref.~\onlinecite{Santoro29032002}). 

\subsection{Path Integral Monte Carlo}

The simplest way to sample $Z$ is to perform Metropolis Monte Carlo moves on this extended system; the Metropolis weight being the integrand of Eq.~\ref{eqZ}.
Even in the continuum case, updates can be \emph{local} or \emph{global}.
The simplest local update consists in moving only one coordinate replica $x_k' \rightarrow x_k + \delta z $ and accept/reject the move accordingly to the Metropolis algorithm.
 $z$ is a uniform random number in the range $[-1,1]$ and $\delta$ is tuned in order to optimize the autocorrelation times.
 This is the kind of local update used in the main text. One MC sweep consists of $P$ local attempts.
 Clearly, many more global updates exist in which several replicas are moved at the same time. Global moves are extremely more efficient than the local ones for realistic many particles system \cite{RevModPhys.67.279,tuckermann_staging1993}.
 
 \subsection{Path Integral Molecular Dynamics}
 
 Employing Molecular dynamics (MD) to sample the finite temperature canonical distribution of the ring polymer\cite{tuckermann_staging1993} is another way to perform  global updates, which has no counterpart in spin models.
 In this case the forces  are used to drive the sampling.
 Among all the possible MD integrator scheme we choose the Langevin equation of motions, because  it represents a simple thermostat for sampling canonical distribution and allows ergodic sampling\cite{mazzola_unexpectedly_2014}.
Adding the conjugate momenta $\pi$ to the $x$ coordinates, the ring polymer Hamiltonian reads
 \begin{align*}
 H_{cl} = \sum_{k=1}^P \left(   {  \pi_k^2 \over 2  } + {m(PT)^2 \over 2}  (x_k - x_{k+1})^2 + {1\over PT}  V(x_k) \right) 
 \end{align*}
 In order to sample the equilibrium distribution $e^{-H_{cl}/PT}$ we integrate the (discretized) second order Langevin equation of motion
 \begin{eqnarray}
  \pi_k' &=&   ( 1 - \delta \gamma) \pi_k  + \delta /m ~ f_k + \sqrt{2\gamma T \delta / m} ~ \eta \\
  x_k' &=& x_k + \pi_k \delta
 \end{eqnarray}
 where $\delta$ is the integration time step,  $\eta$ is a Gaussian distributed random number, and $f_k$ includes all the forces acting on the $k$-th replicas,
 \begin{equation}
 f_k = -{\partial V(x_k) \over \partial x_k} /P + PT^2m (x_{k+1} -2x_k + x_{k-1} )
\end{equation} 
In this equation $\gamma$ is a parameter which has to be tuned in order to minimize autocorrelation times and -in general- can also be position dependent.
All the (collective) moves are accepted at the cost of introducing a time-step discretization error, which can be systematically removed in the limit $\delta \rightarrow 0$.
This is the kind of global update used in the main text.

\subsection{Model, instanton detection and results}

The simple onedimensional model considered in this work is given by the double well potential $V(x) = \lambda x^4 - x^2 $. 
The smaller the $\lambda$ parameter the larger is the distance $d$ between the two wells, which goes as $d = 2 x_{min} = 2 \sqrt{1/ 2\lambda}$. The energy barrier also grows with decreasing $\lambda$, as 
$\Delta V = 1/4 \lambda$.
Therefore the control parameter $\lambda$ in this toy model shares the same features as the system size $L$ for the transverse field Ising model, defined in the main text.
Accurate energy gaps $\Delta$, between the ground and the first excited states, as a function of $\lambda$  are calculated in Ref.~\cite{PhysRevD.18.4767} for the Hamiltonian $H = p^2 + \lambda x^4 - x^2$, which can be used for direct comparison if we set the mass $m=1/2$ in the above equations. 
In this work  $\lambda$ spans between $0.05$ and $0.20$ which can provide a broad range of $1/\Delta^2(\lambda)$ values, varying over   5 orders of magnitude.
We choose the temperature $T \ll \Delta V $ in order to rule out thermally activated transitions. We set $T=0.05$ and $P=64$. 
We start the simulations with all the replicas located at the bottom of the right well, i.e. $x_k = x_{min}$. We identify the tunneling time in the following way:
when the $25\%$ of the replicas exceed the coordinate $-x_{min}/2$ we stop the simulation and record this passage time expressed as the number of MC sweeps (in the case of local updates with PIMC) or  MD integration steps (for PIMD).

  \begin{figure}[t]
 \includegraphics[width=1\columnwidth]{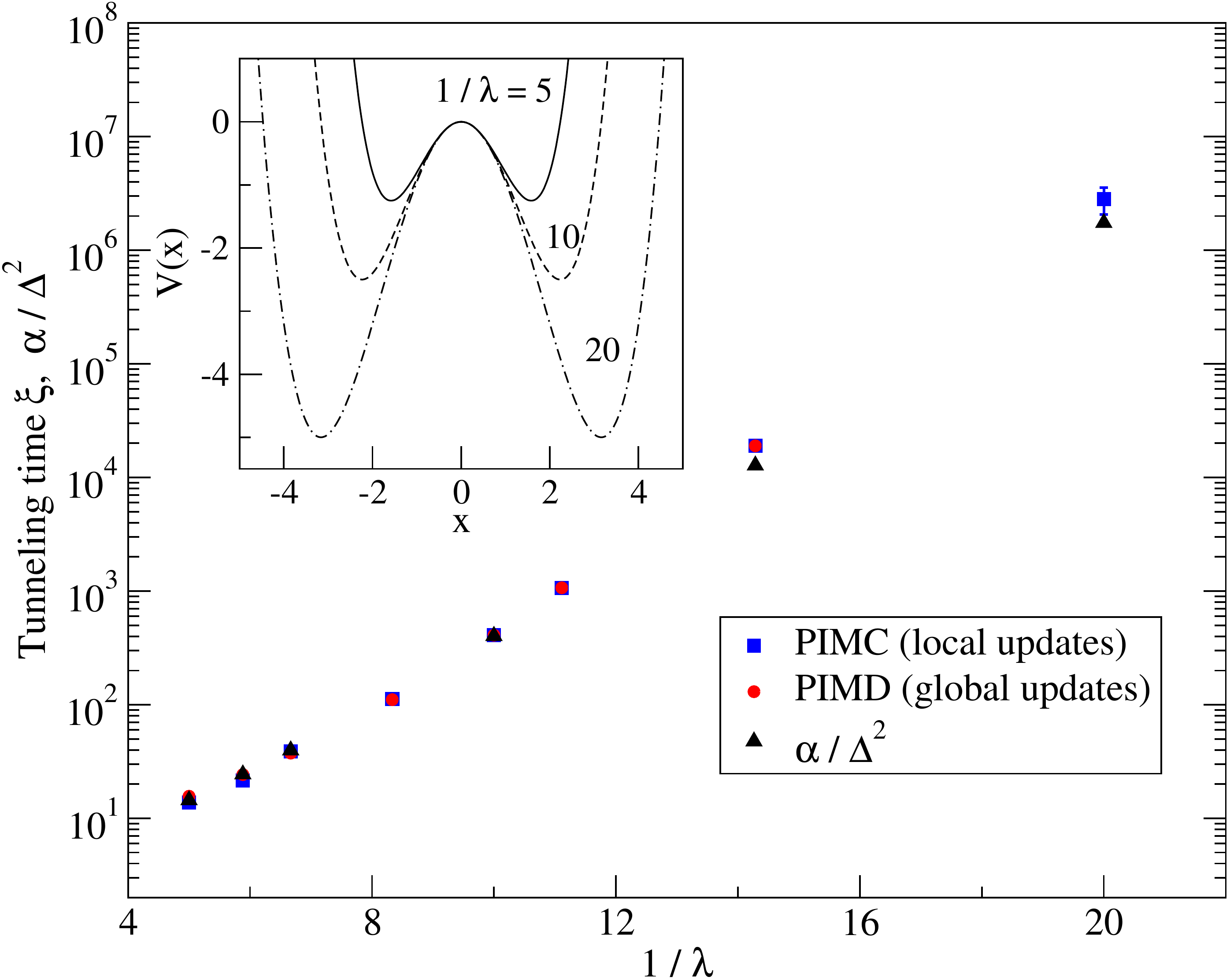}
 \caption{ (color online). Average QMC tunneling time as a function of $1/\lambda$  at $\beta=20$, corresponding to a temperature always much lower than the barrier height. Simulations are performed with two different types of updates: local updates (PIMC) and global updates, using path integral molecular dynamics (PIMD). Finally, we show $\alpha/\Delta^2 (\lambda)$, from Ref.~\onlinecite{PhysRevD.18.4767} scaled by a constant $\alpha$ to demonstrate once more the identical scaling for tunneling times varying over five order of magnitudes. The inset shows the double well potential $V(x)$ for three different values of $1/\lambda$=5, 10, 20. 
 }
 \label{fig:toy}
 \end{figure}
 
 \section{Tunneling in ferromagnetic p-spin model}
 
We first summarize the theory of tunneling in p-spin model closely following the theory developed in \cite{KostyaVadim2015}.
Consider the quantum Hamiltonian for an $L$-qubit system corresponding to a ferromagnetic p-spin model in a transverse  field
\begin{align}
 \hat H = -\Gamma\sum_{j=1}^{L}\hat\sigma_{x,\,j} -L g\left( \frac{1}{L}\sum_{j=1}^{L}\hat\sigma_{z,\, j}\right)\;. \label{eq:H-sigma}
\end{align}
Here  $\Gamma$ is the strength of the transverse field and the  function  $\g$ is the mean-field potential energy of the spin system. We will consider the case  where the function $g$ admits a metastable state and will be interested in the  tunneling decay of this state at zero temperature.
A well known example is given by the Curie-Weiss model where the function $\g$ takes the form
\begin{align}
 \g(m) = \half  m^2 +h  m\label{eq:g}
\end{align}
where $h$ is a local field.   

 The Hamiltonian  (\ref{eq:H-sigma}) is symmetric with respect to permutation of individual spins. To understand the structure of the eigenstates we use the   total spin operators 
\begin{align}
 \hat S_x = \frac{1}{2}\sum_{j=1}^{L}\hat\sigma^x_j\,,\quad
 \hat S_y = \frac{1}{2}\sum_{j=1}^{L}\hat\sigma^y_j\,,\quad
 \hat S_z = \frac{1}{2}\sum_{j=1}^{L}\hat\sigma^z_j.\label{eq:Sxyz}
\end{align}
The system Hamiltonian can be written in terms of these operators:
\begin{equation}
\hat H=-2\Gamma \hat S_x- L g(2 \hat S_z/L)\label{eq:H}
\end{equation}
We are interested in the eigenstates of the Hamiltonian $H |\Psi \rangle=E |\Psi \rangle$. 
They   correspond to certain values of the  total spin operator 
\begin{align}
 \hat S^2 = \hat S_x^2 + \hat S_y^2 + \hat S_z^2, \label{eq:S2}
\end{align}
which can take the following discrete values
\begin{align}
 S = \frac{L}{2} - K\,,\quad K = 0, 1, 2, \ldots, \Big\lfloor \frac{L}{2} \Big\rfloor \label{eq:S}
\end{align}
The number of distinct irreducible subspaces of a given total spin $S$ is
\begin{align}
 {\mathcal N}(L, S) = \binom{L}{K} - \binom{L}{K-1}\label{eq:sN}
\end{align}
each subspace has dimension $2S+1 = L-2K+1$.
We will expand the eigenstates of $H$ in the basis of the operators $S_z$ and $S^2$
\begin{equation}
|\Psi\rangle=\sum_{S=-L/2}^{L/2}\sum_{\gamma=1}^{ \sN(L, S)} \sum_{M=-S}^{S}C_{M}^{S,\gamma}|M,S,\gamma\rangle\label{eq:Psi}
\end{equation}
where 
\begin{align}
S^2|M,S,\gamma\rangle &= S(S+1)|M,S,\gamma\rangle,\\ S_z |M,S,\gamma\rangle &= M |M,S,\gamma\rangle\label{eq:basis}
\end{align}
and the  coefficients $C_{M}^{S,\gamma}$ obey  the stationary \Sch equation
\begin{multline}
-\Gamma \sqrt{(S+M)(S-M+1)} C_{M-1}\\-\Gamma \sqrt{(S+M+1)(S-M)} C_{M+1}\\-L g(2M/L) C_M=E C_M
\end{multline}
(here we omitted the indices $\gamma$ and $S$.) 

In what follows we will study the limit  $L \gg 1$ and assume  that for the case of interest $S={\cal O}(L)$. We employ the discrete Wentzel-Kramers-Brillouin method  developed in  Ref.~\cite{Garg-wkb}  applied to a mean-field model of a composite   large spin~\cite{Semerjian-wkb,KostyaVadim2015}.   	

To the leading order in the  parameter $S$ the  stationary \Sch     equation  can  be  written in the form
\begin{align*}
-\left (2\Gamma \sqrt{S^2-M^2}\cos \hat p+L f(2M/L) \right)
  \,C_{M}&=E\, C_{M},\\ p&=-i \frac{\partial }{\partial M}\;, 
\end{align*}
where we omitted the indices $S$ and $\gamma$ for brevity.  Following the discrete WKB approach we write  the wave function amplitudes $C_M$  as follows
\begin{align}
C_M&=\frac{1}{\sqrt{v(M,E)}} \exp\left(i {\cal A}(M,E)
     \right)\label{eq:Cm} \\
{\cal A}(M,E)&=\int^{M} dM'\, p(M',E)\label{eq:Sc}
\end{align}
where the  momentum $p(M,E)$ obeys the classical Hamilton-Jacobi equation
\begin{equation}
H(M,p)=-2\Gamma \sqrt{S^2-M^2}\cos p-L g(2M/L) =E\label{eq:Hcl}\;.
\end{equation} 
We note that in the classically-allowed region, in the case relevant to us, $ 0\leq \cos p \leq 1$ and therefore
\begin{equation}
2\Gamma \sqrt{S^2-M^2} \geq -(L g(2M/L)+E)\geq 0 \label{eq:al}\;.
\end{equation}

We now introduce the rescaled quantities
\begin{align}
m&=\frac{L}{2} M & E &= L e \nonumber \\ -i{\cal A}&=\frac{L}{2}a & S &=
                                                               \frac{\ell
                                                               L}{2} \quad\ell \in [0,\,1]\label{eq:res}\;.
\end{align}
We note that with the above rescaling we must also introduce a new time variable  $s$. In  order to satisfy  the condition 
\begin{equation}
 \frac{dM}{d\tau} = L \frac{dm}{ds}
\end{equation}
we must set 
\begin{equation}
s=2\tau 
\label{eq:s}
\end{equation}
Using (\ref{eq:res})  we write 
\begin{equation}
e(m,k,\ell)=-\Gamma \sqrt{\ell^2-m^2}\cosh k  - g(m) \label{eq:Ham-e}
\end{equation}

\begin{figure*}[t]
\begin{minipage}{.45\textwidth}
     \includegraphics[width=0.9\textwidth]{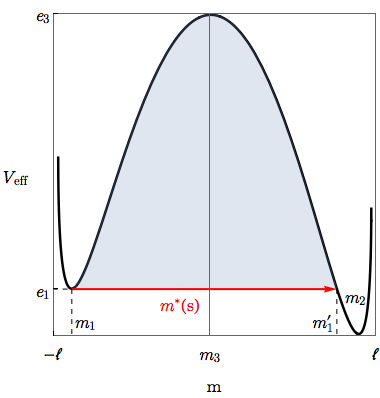}
\end{minipage}
\begin{minipage}{.45\textwidth}
       \includegraphics[width=0.9\textwidth]{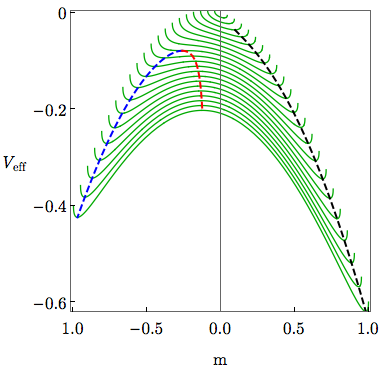}
\end{minipage}
    \caption{(Left) Plot   of the  effective potential
      (\ref{eq:Veff}) $V_{\rm eff}(m,\ell)$ {\it vs} $m$ for the
      Curie-Weiss model for $\ell=1$, $\Gamma=0.4$ and $h=0.015$. Red
      line depicts the tunneling path (instanton) at zero
      temperature. Barrier energies above the instanton energy are
      shown with blue filling. (Right) Plots of the effective
      potential for the Curie-Weiss model at different values of
      $\ell$ shown with green lines.  Dashed blue, red and black lines
      are $\{m_k(\ell),e_k(\ell)\}_{\ell=\ell_c}^{\ell=1}$  for
      $k=1,2,3$ corresponding to the local minimum, maximum  and global
      minimum respectively.   $\ell$ changes in the range from
      $\ell_c$ to  1 where  $\ell_c=(h^{2/3}+\Gamma^{2/3})^{3/2}$ is
      the smallest value of $\ell$ beyond which the effective
      potential is monostable. The domain of the effective potential
      for each $\ell$ is    $m\in(-1,1)$. We see by comparison of figures in left and right panels that  for a given $\ell$ the instanton turning point at the barrier exit satisfies the condition $m_1^\prime \leq \ell$ where the equality is reached only for $\ell=1$.    In all the plots $h=0.1$ and $\Gamma=0.21$. }
 \label{fig:Veff}
\end{figure*}

 We introduce the effective potential
\begin{align}
V_{\rm eff}(m,\ell) =e(m,0,\ell)= -\Gamma \sqrt{\ell^2-m^2} -\g(m) \label{eq:Veff}
\end{align}
 This potential corresponds to the energies for the turning points $m$ of the classical trajectories of the system with the Hamiltonian function (\ref{eq:Ham-e}).
 The   potential at different values of $\ell$ is  depicted in the right panel of Fig.~\ref{fig:Veff}.
 It  has a double well shape in certain  parameter range of $\ell$. We
 denote the extreme points of the potential as
 $\{m_k(\ell),e_k(\ell)\}$, which for k=1,2,3 correspond to the local
minimum, maximum and global minimum  respectively (cf. right panel of Fig.~\ref{fig:Veff}).

The  imaginary  momentum in the classically forbidden region under the barrier can be written in the form 
\begin{equation}
 k(e,m,\ell)={\rm arcsinh}
 \frac{\sqrt{(e+g(m))^2-\Gamma^2(\ell^2-m^2)}}{\Gamma\sqrt{\ell^2-m^2}}   \label{eq:kr}\;.
\end{equation}
We also note that in the classically forbidden region
\begin{equation}
 -(e+g(m)) > \Gamma \sqrt{\ell^2-m^2} >0 
 \end{equation}
The usual canonical relation between velocity and momentum has the form 
\begin{equation}
\frac{\partial k(e,m, \ell)}{\partial e}=-\frac{1}{\nu (e,m,\ell)},\quad  \nu(e,m,\ell) \equiv \frac{dm}{ds}\label{eq:k-nu}
\end{equation}
(the minus sign here occurs because both the velocity and the momentum are purely imaginary under the barrier).
From here we get
\begin{align*}
\nu(e,m,\ell)\equiv \nu(e,m,\ell)=\sqrt{(e+g(m))^2-\Gamma^2(\ell^2-m^2)}\label{eq:nu}
\end{align*}
This expression  also can be obtained from the  Hamiltonian equation for coordinate  $m$.  We  write
\begin{align}
\frac{d m}{ds}&=-\frac{\partial e}{\partial k}\\&=\Gamma\sqrt{\ell^2-m^2}\sinh k\\& = 
\sqrt{(e+g(m))^2-\Gamma^2(\ell^2-m^2)}\label{eq:nuH}
\end{align}

For values of $\ell$ where metastability exists (cf. Fig.~\ref{fig:Veff}) we write the tunneling  amplitude $\Delta=\Delta(e,\ell)$  to logarithmic accuracy     
\begin{align}
\Delta(e,\ell) &\propto  \exp\left(-\frac{L}{2}  a(e,\ell)  \right), \label{eq:Ctunn} \\
 a(e,\ell) &=\int_{m_1(e,\ell)}^{m_1^\prime(e,\ell )}  k(e,m,\ell)  dm \label{eq:a}\;.
\end{align}
Here $m_1$  and  $m_1^\prime$ are turning points of the instanton trajectory on different sides of the barrier that satisfy the condition 
\begin{equation}
e(m_i(e,\ell),k=0,\ell)=e,\quad i=1,2  \label{eq:qi}
\end{equation}
 (the instanton trajectory at zero temperature has  the energy of the local minimum).

\subsubsection{Curie-Weiss model with zero bias}

 The Hamiltonian for the unbiased Curie-Weiss model corresponds to (\ref{eq:H-sigma}) and (\ref{eq:g}) with $h=0$
\begin{equation}
H=-2\Gamma S_x-\frac{2 S_z^2}{L}  \label{eq:H-0bias}\;.
\end{equation}
At zero temperature the tunneling occurs exclusively within the $S=L/2$ subspace of the maximum total spin.  The analytical expression for the tunneling splitting $\Delta_{10}$ of the ground state energy can be derived in the asymptotical limit of large number of qubits   $L\gg 1$ and has a form derived by a number of authors in the past (see e.g. \cite{Garg-wkb} )
\begin{align}
\Delta_{10}^{\rm analyt} & =b(\Gamma) e^{- L c(\Gamma)}\label{eq:Delta}\\
b(\Gamma) &= \left(\frac{32 L}{\pi}\right)^{1/2}     \frac{(1-\Gamma) ^{5/4}}{1+\sqrt{1-\Gamma^2}}   \label{eq:b} \\
c(\Gamma) &= \frac{1}{2}\log\left(\frac{1+\sqrt{1-\Gamma^2}}{1-\sqrt{1-\Gamma^2}}\right)-\sqrt{1-\Gamma^2}\label{eq:c}\;.
\end{align}
Because we will be comparing the (doubled) scaling exponent $c(\Gamma)$  (\ref{eq:c}) with the scaling exponent  for the QMC tunneling rate obtained from the numerical simulations of the problems with limited values of $L$ it is important to understand the finite size effect coming from  the prefactor $b$ in (\ref{eq:b}).

\begin{figure*}[t]
\begin{minipage}{.48\textwidth}
    \includegraphics[width=3.2in]{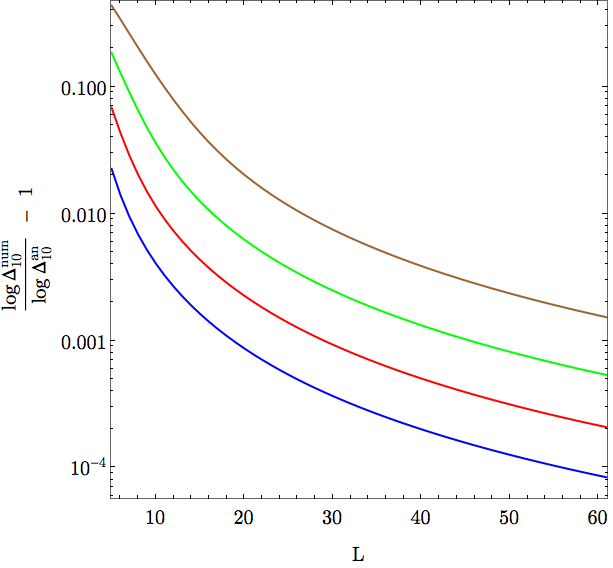}  
\end{minipage}
\begin{minipage}{.48\textwidth}
        \includegraphics[width=3.2in]{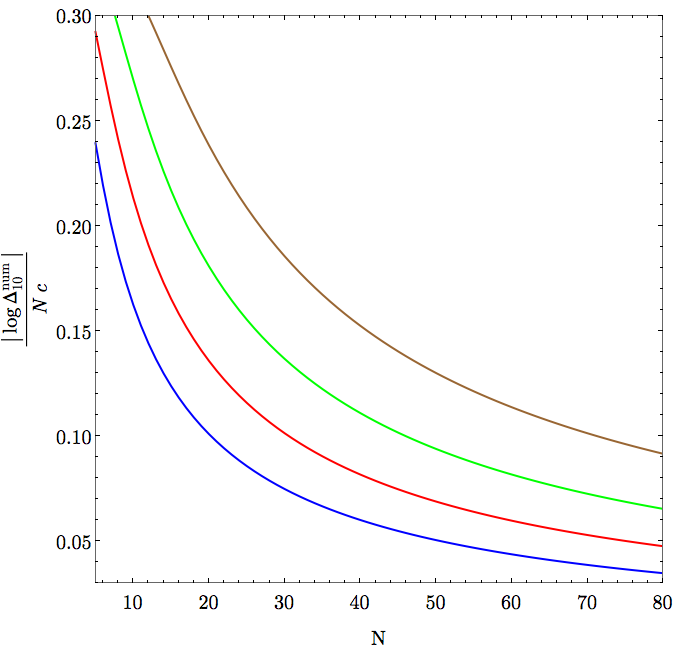}
\end{minipage}
    \caption{The left plot shows the size $L$ dependence of the ratio of the quantity $|\log \Delta_{10}|/L$  computed via  numerical diagonalization  of $H$ (\ref{eq:H-0bias}) to the same quantity given by  WKB (\ref{eq:Delta}).  
  Different curves correspond to $\Gamma$ =0.3, 0.4, 0.5, 0.6. They are depicted with  blue, red, green and brown colors, respectively.  The deviation of the ratio from unity is ${\cal O}(1/L^2)$.
   The right plot shows the size $L$ dependence of the ratio of the exponents  of the energy splitting  $\Delta_{10}$ computed numerically and  the analytical exponent c from Eq.~\eqref{eq:c}. Different curves correspond to $\Gamma$ =0.3, 0.4, 0.5, 0.6. They are depicted with  blue, red, green and brown colors, respectively.  The deviation of the ratio from unity is ${\cal O}(1/L)$.}
 \label{fig:gap0}
\end{figure*}

The left panel of Fig.~\ref{fig:gap0} shows the $L$-dependence of the ratio  of the logarithm of the gap computed  via  the above analytical expression to that computed  via   the  numerical diagonalization of  the Hamiltonian (\ref{eq:H-0bias}). Similarly, the right panel of Fig.~\ref{fig:gap0} shows the comparison between the scaling exponent $c$ (\ref{eq:c})  and the quantity $|\log \Delta_{10}/L|$  computed via  numerical diagonalization  of $H$.  Note that in the later case the convergence with $L$ is ${\cal O}(1/L)$ while in the former it is ${\cal O}(1/L^2)$.

\begin{figure}[ht]
\centering
\includegraphics[width=1\columnwidth]{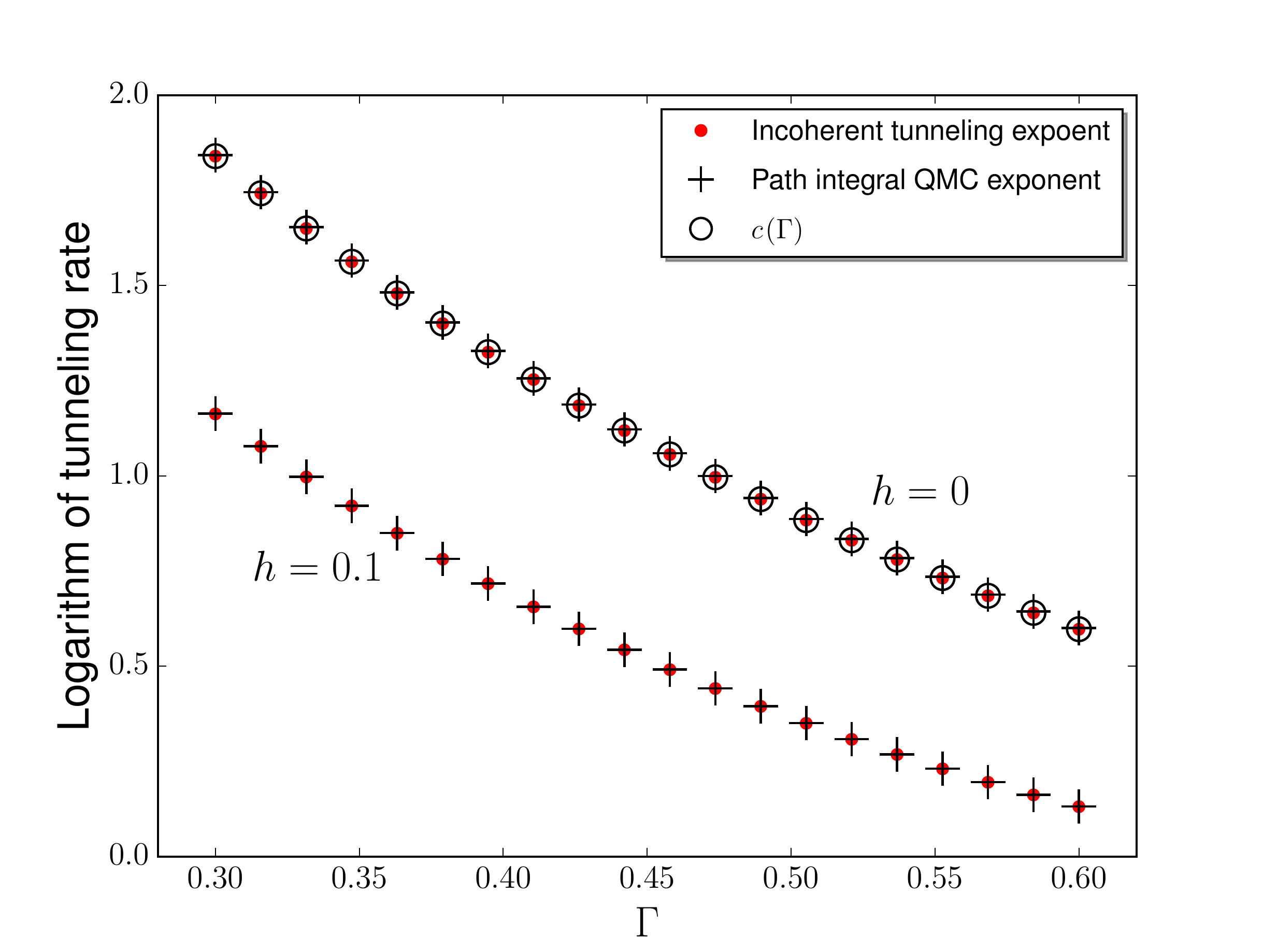}
   \caption{Plot show the dependence on $\Gamma$ of the exponents  of the tunneling decay rate for the metastable state for the Curie-Weiss model at zero temperature. Red points correspond to the  
    incoherent tunneling rate  given by $a(e_1,1)$ (\ref{eq:Ctunn}) corresponding to the tunneling with maximum total spin. Circles correspond to  $2c(\Gamma) $ (\ref{eq:c}) for the case of $h=0$. Crosses correspond to the exponent of the path integral QMC tunneling rate estimated using Kramers formulae.
       \label{fig:comp}}
\end{figure}

Finally, Fig.~\ref{fig:comp} shows the dependence on $\Gamma$ of the exponent  of the tunneling decay rate of the  metastable  state given by (\ref{eq:Ctunn}) (for zero basis the result is also computed by  using (\ref{eq:c}) ).

\subsection{Tunneling in the Grover problem \label{eq:Grover}}

The Grover search problem can be recast in the adiabatic setting using the following Hamiltonian (see \cite{farhi_quantum_2000})
\begin{align}
 \hat H &= -\Gamma\sum_{j=1}^{L}\hat\sigma^x_j -L
  |\pmb{\uparrow}\rangle\langle\pmb{\uparrow}|, \\
 |\pmb{\uparrow}\rangle &= |\uparrow\rangle_1\otimes  |\uparrow\rangle_2\cdots \otimes  |\uparrow\rangle_L
  \label{eq:HGrover}
\end{align}
This Hamiltonian can be rewritten in terms of the large spin operators (\ref{eq:Sxyz}) in the form (\ref{eq:H}) with the function $g(m)$ 
\begin{align}
g(m) &=\delta\left(\frac{L(1-m)}{2}\right),\\ m&=-1,-1+\frac{2}{L},\ldots,1-\frac{2}{L},1
\end{align}
where $\delta(m)$ is  the Kronnecker delta.

At $\Gamma>1$ the ground state of this Hamiltonian   is
$|\pmb{\rightarrow}\rangle$,  corresponding to all spins along the $x$ axis. The ground state energy is $-L \,\Gamma$.  At $\Gamma<1$ the ground state is
$|\pmb{\uparrow}\rangle$ with energy $-L$.  At $\Gamma\approx 1$ the avoided crossing occurs between the energies with the size of the energy splitting being
\begin{equation}
\Delta\sim \langle\pmb{\rightarrow}|\pmb{\uparrow}\rangle=2^{-L/2}\label{eq:G}
\end{equation}
We now recast this result using the WKB approach described above. For
$m<1$ the equation (\ref{eq:Ham-e}) takes the form
$e(m,k,\ell)=-\Gamma \sqrt{\ell^2-m^2}\cosh k$ . The  corresponding
effective potential is
\begin{align}
V_{\rm eff}(m,\ell)=\left [
\begin{array}{cc}
-\Gamma \sqrt{\ell^2-m^2},   & m< \ell   \\
 -\delta_{\ell,1}, & m=\ell 
\end{array} \right.
  \label{eq:Veff2}
\end{align}
In complete analogy to the discussion in the general case presented in
the previous section the ground state wavefunction is localized at $m=0$ and decays exponentially away from this point. The tunneling 
path moves under the barrier and ends at the point $m=\ell$. It is
immediately clear that the tunneling path can only reach the target
state for $\ell=1$. In fact this  also immediately follows from inspection of the Fig.~\ref{fig:Veff} (right panel). While this figure is plotted for a different $g(m)$ the common feature is that  the tunneling  path terminates at $m_1^\prime<= \ell$ where the equality is reached only for $\ell=1$.

At the avoiding crossing $\Gamma=1$ we solve the equation (\ref{eq:nuH})  for the instanton trajectory $m=m^*(s)$ for  $\ell=1$ and get
\begin{equation}
m^*(s)=e^{-s},\quad -\infty< s< 0 ,\quad (\ell=1,\,\,\Gamma=1)
\end{equation}
This trajectory approaches the solution state at $m=1$ as
$s\rightarrow 0$. The momentum (\ref{eq:kr}) at the instanton trajectory under the barrier equals
\begin{equation}
\kappa  (m)=\frac{\text{\rm arcsinh}( m)}{\sqrt{1-m^2}}
\end{equation}
With that the action (\ref{eq:a}) along the instanton is
\begin{equation}
a=\log 2
\end{equation}
Using (\ref{eq:Ctunn}) we recover the well-known quadratic speedup (\ref{eq:G}).

 \section{Tunneling in path-integral QMC for ferromagnetic p-spin model}
To describe the stochastic QMC process  that  samples  paths  over imaginary time in a spin system with the Hamiltonian  (\ref{eq:H-sigma}) we 
 define  the state vector
\begin{equation}
{\underline \sigma}(\tau)=\{   \sigma_{1}(\tau),\ldots,\sigma_{N}(\tau)\}   \label{eq:us}
\end{equation}
In path integral QMC the trajectories satisfy periodic boundary conditions
\begin{equation}
\quad {\underline \sigma}(0)= {\underline \sigma}(\beta)\label{eq:pbc}
\end{equation}
In the PIGS version of QMC one  uses open boundary conditions.
There exists an order parameter corresponding  to the z-component of the  total magnetization 
\begin{equation}
m[\s(\tau)]=\frac{1}{L}\sum_{i=1}^{L} \sigma_{i}(\tau)\nonumber
\end{equation}
The probability functional for a realization of the trajectory ${\underline \sigma}(\tau)$ equals
\begin{align}
P[\s(\tau)]&=\frac{1}{Z} e^{ -{\cal F}[\s(\tau)] }\nonumber \\
{\cal F}[\s(\tau)] &=  \int_{0}^{\beta}g\left(m[\s(\tau)]\right) d\tau -\sum_{j=1}^{N} \log w[\sigma_{j}(\tau)]\nonumber
\end{align}
Above  $w[\sigma(\tau)]$ is a  single-spin imaginary time propagator  evaluated along  the trajectory $ \sigma(\tau)$ (see e.g. \cite{Semerjian-cavity} for details). 
Marginalizing over the path configurations that corresponds to a given $m(\tau)\equiv m[\s(\tau)] $  the probability functional  can be written in the form (cf. \cite{Semerjian-wkb})
\begin{equation}
\mathbb{P}[m(\tau),\lambda(\tau)]=\frac{1}{\mathbb{Z}}\,  e^{-L \,\mathbb{F}[m(\tau),\lambda(\tau)]}
\end{equation}
\begin{align}
\mathbb{F} =\int_{-\beta/2}^{\beta/2}d\tau \left(\lambda(\tau) m(\tau)-g[m(\tau)]\right)-\log D[\lambda(\tau)],\label{eq:SS}
\end{align}
where
\begin{align}
D[\lambda(\tau)] &={\rm Tr} K[\beta/2,-\beta/2; \,\lambda(\tau)]  \label{eq:D}\\
K[\beta/2,-\beta/2; \,\lambda(\tau)] &= T_+ \exp \left(-\int_{-\beta/2}^{\beta/2} H_0(\tau) d\tau\,\label{K} \right)\\
H_0(\tau) &=-\Gamma \sigma_x  -\lambda(\tau)\sigma_z \label{eq:H0}
\end{align}
The partition function $\mathbb{Z}$ is defined as usual as a normalization factor in the  probability functional. In the above $\lambda(\tau)$  is a classical \lq\lq gauge" field associated with $m(\tau)$. 

 This functional has \lq\lq trivial" extremal points  $m(\tau)=\bar m_k$ 
\begin{equation}
\frac{\delta \mathbb{F}}{\delta m(\tau)}=0,\quad  \frac{\delta \mathbb{F}}{\delta \lambda(\tau)}=0,\quad {\rm for}\quad  m(\tau) = \bar m_k,\quad \lambda(\tau)=\bar\lambda_k\
\end{equation}
corresponding to the  extrema of the mean-field free energy function 
\begin{align}
f(m,\lambda) &=\lambda m-g(m) -\frac{1}{\beta} \log 2\cosh(\beta\sqrt{\lambda^2+\Gamma^2}) \nonumber \\
&=\lambda m-g(m) -\sqrt{\lambda^2+\Gamma^2}) \quad (\beta\gg 1)\label{eq:free}
\end{align}
\begin{equation}
\bar \lambda_k=\frac{\partial g(\bar m_k)}{\partial m},\quad \bar m_k=\frac{\partial f(\bar m_k,\bar \lambda_k)}{\partial m}=0
\end{equation}
For the sake of bookkeeping we assume that $\bar m_1$ and $\bar m_2$ correspond to  local and global minima respectively,
\begin{equation}
f_k=f(\bar m_k,\bar \lambda_k),\quad  f_1>f_2
\end{equation}
The free energy functional values at the minima equal
\begin{equation}
\mathbb{F}_k \equiv \beta f_k,\quad k=1,2 \label{eq:Fk}
\end{equation}

During  QMC a stochastic evolution of the system state (\ref{eq:us}) corresponds to a stochastic evolution   of  $m(\tau,t)$, $\lambda(\tau,t)$ where $t$ is Monte Carlo time.
We initialize the system state at the vicinity of the local minimum of the free energy functional $\mathbb{F}$ 
\[m(\tau,0)\approx \bar m_1,\quad \lambda(\tau,0)\approx \bar \lambda_1\]
Most of the time $\{m(\tau,t),\lambda(\tau,t)\}$ fluctuates near the  local minimum. Escape from this minimum can be described in terms of the classic Kramers  escape theory as was illustrated in the paper for the case of QMC with continuous variables. To logarithmic accuracy the escape rate $W$ is given by the Boltzmann factor
\begin{equation}
W\propto e^{  -(\,\mathbb{F}[m^*(\tau),\lambda^*(\tau)]-\beta f_1)}
\end{equation}
where we assumed that the Monte Carlo temperature equals 1. Here $m^*(\tau),\lambda^*(\tau)$ correspond to a  nontrivial saddle point of the free energy functional satisfying $\delta \mathbb{F}=0$
\begin{equation}
\begin{array}{ccc}
 \frac{\delta \mathbb{F}}{\delta m(\tau)}=0 &\Rightarrow   &  \lambda(\tau)=\frac{\partial g(m(\tau)}{\partial m} \\
  &   &   \\
\frac{\delta \mathbb{F}}{\delta \lambda(\tau)}=0  & \Rightarrow   &   m(\tau)=\frac{1}{D[\lambda(\tau)]}\frac{\partial D[\lambda(\tau)]}{\partial \lambda(\tau)}\
\end{array}
\label{eq:dD}
\end{equation}
Using  (\ref{eq:D}),(\ref{K}) and (\ref{eq:H0}) we obtain
\begin{equation}
m(\tau)=\frac{{\rm Tr}[K(\beta/2,\tau)\sigma_z K(\tau,-\beta/2)]}{D}\label{eq:sce}
\end{equation}
where we dropped the argument $\lambda(\tau)= g^\prime(m(\tau))$ above.
We  introduce a vector
\begin{equation}
{\bf m}(\tau)=
\left(\begin{array}{c}
m_x(\tau) \\ m_y(\tau) \\ m_z(\tau)
 \end{array} \right),\quad m(\tau)\equiv m_z(\tau)
\end{equation}
Then after some transformations, using  (\ref{eq:D}),(\ref{K}), (\ref{eq:H0}) we obtain
\begin{align}
\frac{i}{2}\frac{d{\bf m}}{d\tau}=\frac{\partial {\mathcal H}}{\partial {\bf m}}\times {\bf m(\tau)}\label{eq:dm}
\end{align}
where
\begin{equation}
{\mathcal H}=-\Gamma m_x(\tau)-g(m_z(\tau))\label{eq:cH0}
\end{equation}
The above equations has intervals of motion 
\begin{equation}
{\bf m}\cdot {\bf m}=\ell^2,\quad {\cal H}[{\bf m}(\tau)]=e\label{eq:intm}
\end{equation}
We set
\begin{equation}
m_x=\sqrt{\ell^2-m_z^2} \cosh k,\quad m_y=-i \sqrt{\ell^2-m_z^2}\sinh k\label{eq:mxy}
\end{equation}
where $m_z(\tau)$ satisfies the equation
\begin{equation}
-\Gamma \sqrt{\ell^2-m_z^2}\cosh k-g(m_z)=e\label{eq:eqmc}
\end{equation}
Upon   inspection of the Eqs.~(\ref{eq:eqmc}),(\ref{eq:Ham-e}) and
(\ref{eq:kr}) one can see that up to the choice of the integrals of
motion  $e$ and $\ell$ the equation for $m_z(\tau)$ coincides  with
that for the quantum instanton   described in the previous section.
In the zero-temperature limit of $\beta \gg 1$ the value of $e$ must coincide with the local minimum of $V_{\rm Jeff}(m_z,\ell)$ in $m_z$. Then the choice of $\ell$ must be done in a way that the self-consistent equation (\ref{eq:sce}) is satisfied.

 Detailed calculation will be provided elsewhere \cite{aZSISTH}. One can  show that in the zero temperature limit  $\beta \gg 1$  the value of $\ell=1$ and the  WKB instanton for the tunneling at zero temperature precisely corresponds to the  path  $m^*(\tau)\equiv m_z(\tau)$ that provides the saddle  point of the free-energy functional $\mathbb{F}$ (with time rescaling (\ref{eq:s}) taken into account). Similarly, the action along the WKB instanton trajectory equals to the difference of the values of $\mathbb{F}$ between the saddle point and local minimum
 \begin{align}
 \mathbb{F}[m^*(\tau), \lambda^*(\tau)]-\beta f(\bar m_1,g^\prime(\bar m_1)) = 
 \end{align}
 \vspace{-0.26in}
 \begin{align}
  = \left[ 
 \begin{array}{ccc}2a(e_1,1)  & \hspace{0.4cm} m(-\beta/2)=m(\beta/2)  & \\ a(e_1,1) &   m(-\beta/2)=m_1, &  m(\beta/2)=m_1^\prime
 \end{array}\right.\nonumber
 \end{align}
 Therefore at zero temperature  the exponential scaling with the problem size $L$ of the QMC and quantum tunneling are the same.

\section{ QMC tunneling results for a ferromagnetic Ising chain}

Here we report QMC simulations on a ferromagnetic
Ising  chain in the presence of a weak transverse field $\Gamma$ with Hamiltonian $H=-\Gamma \sum \sigma_j^x - \sum J_{ij} \sigma_i^z \sigma_j^z$. The nearest neighbours  couplings are given by $J_{ij}=\delta_{i,j+1}+\delta_{i,j-1}$.
%

We measure the average tunneling time for different sizes $L$, i.e. the number of sites, as explained in the main text and compare these values with the inverse gap and inverse gap squared obtained with exact diagonalization.

\begin{figure}[htp]
 \includegraphics[width=1\columnwidth]{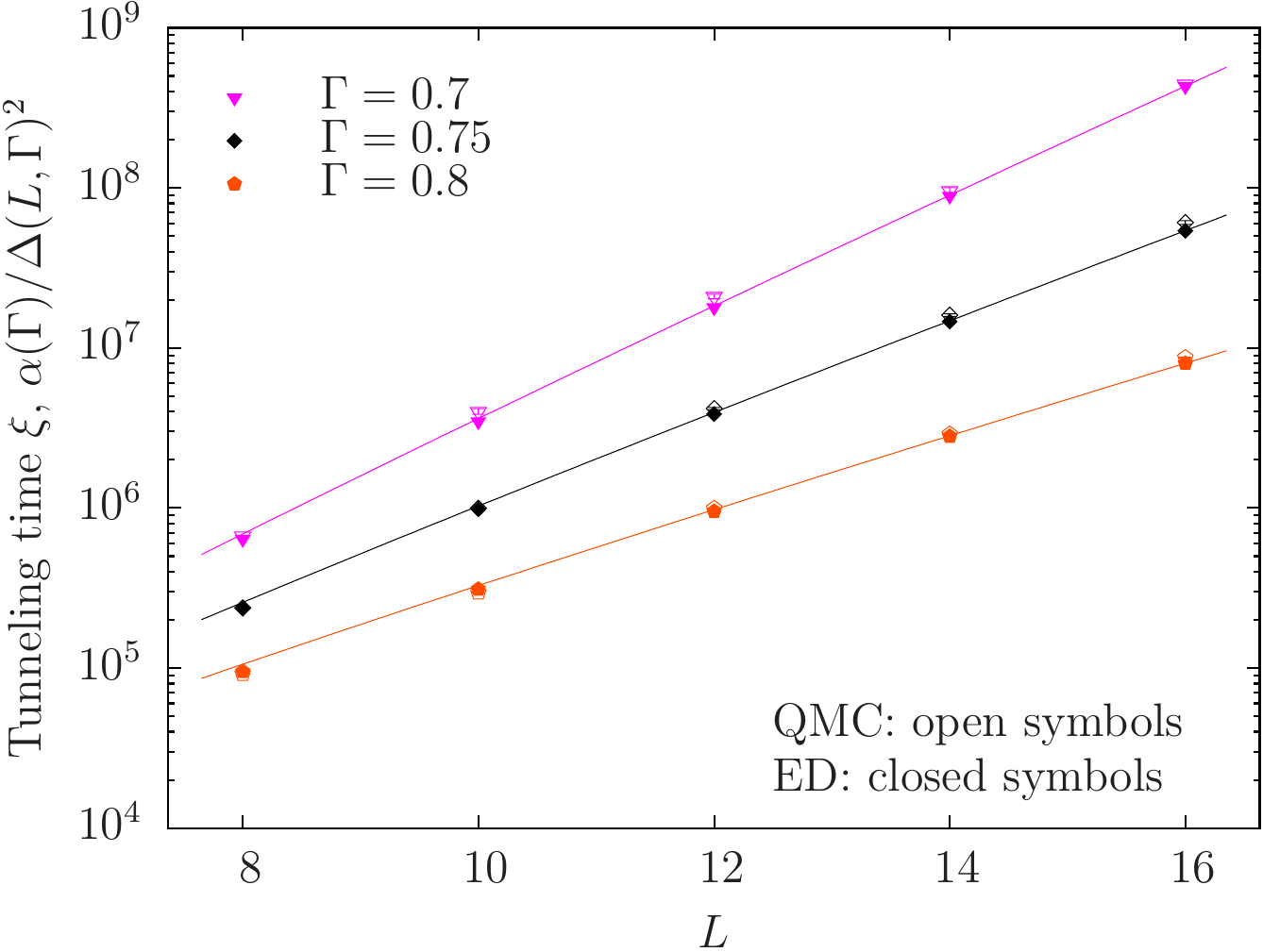}
 \caption{ (color online). Average PIMC tunneling time $\xi$ as a function of system size $L$ (open symbols) at $\beta=24$, for various values of the transverse field $\Gamma$. Exponential fits of the times for $12 \le L \le 16 $ are shown as solid lines. To compare to physical QA we also show $a(\gamma)/\Delta(\Gamma,L)^2$, obtained by exact diagonalization (ED) with solid symbols. Rescaling the latter by $L$-independent constants $\alpha(\Gamma)$ we find identical scaling with system size $L$. 
 }
 \label{f:pbc4}
 \end{figure}

\begin{figure}[htp]
 \includegraphics[width=1\columnwidth]{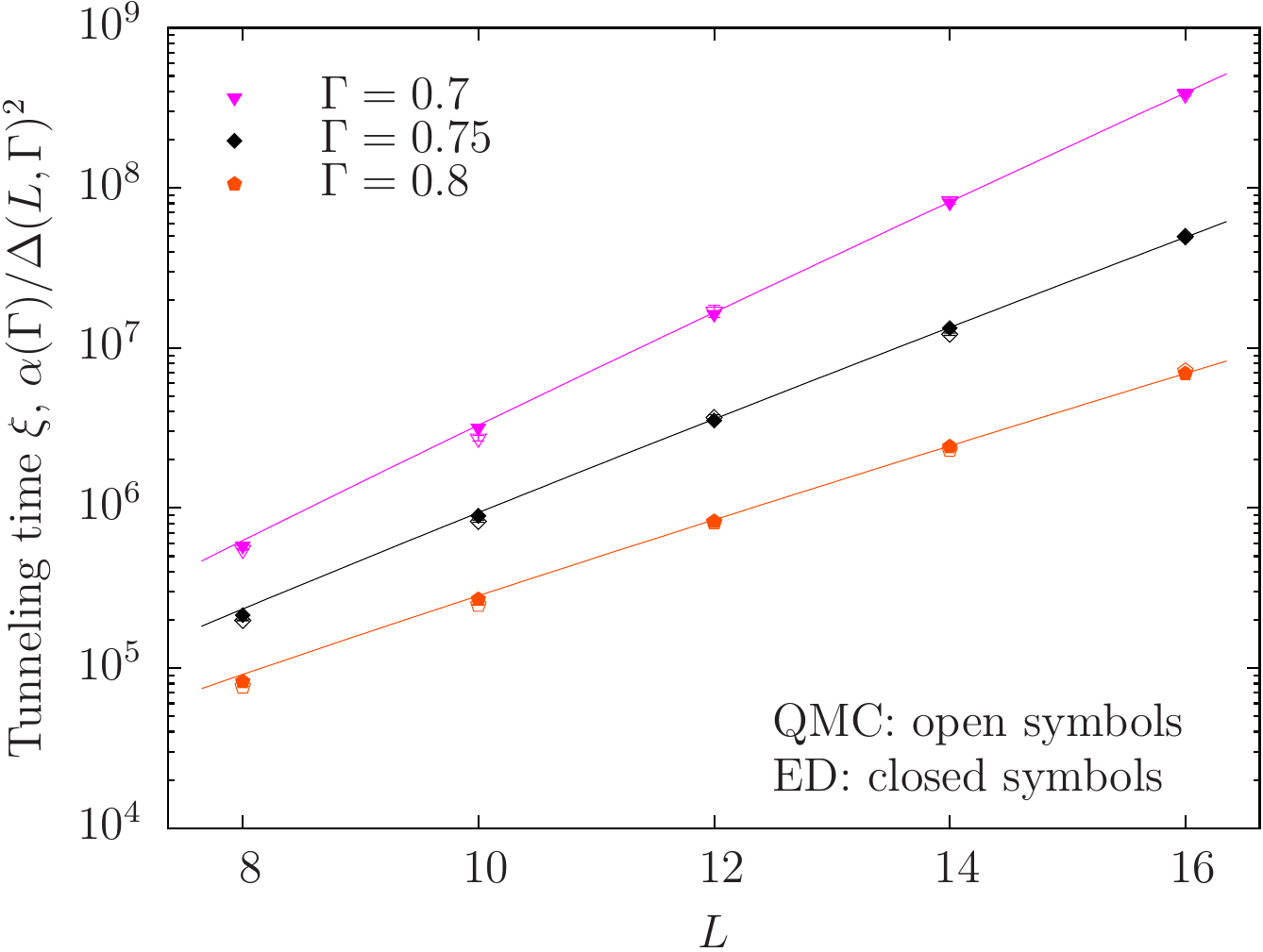}
 \caption{ (color online). Average PIMC tunneling time $\xi$ as a function of system size $L$ (open symbols) at $\beta=20$, for various values of the transverse field $\Gamma$. Exponential fits of the times for $12 \le L \le 16 $ are shown as solid lines. To compare to physical QA we also show $a(\gamma)/\Delta(\Gamma,L)^2$, obtained by exact diagonalization (ED) with solid symbols. Rescaling the latter by $L$-independent constants $\alpha(\Gamma)$ we find identical scaling with system size $L$. 
 }
 \label{f:pbc5}
 \end{figure}

\begin{figure}[htp]
 \includegraphics[width=1\columnwidth]{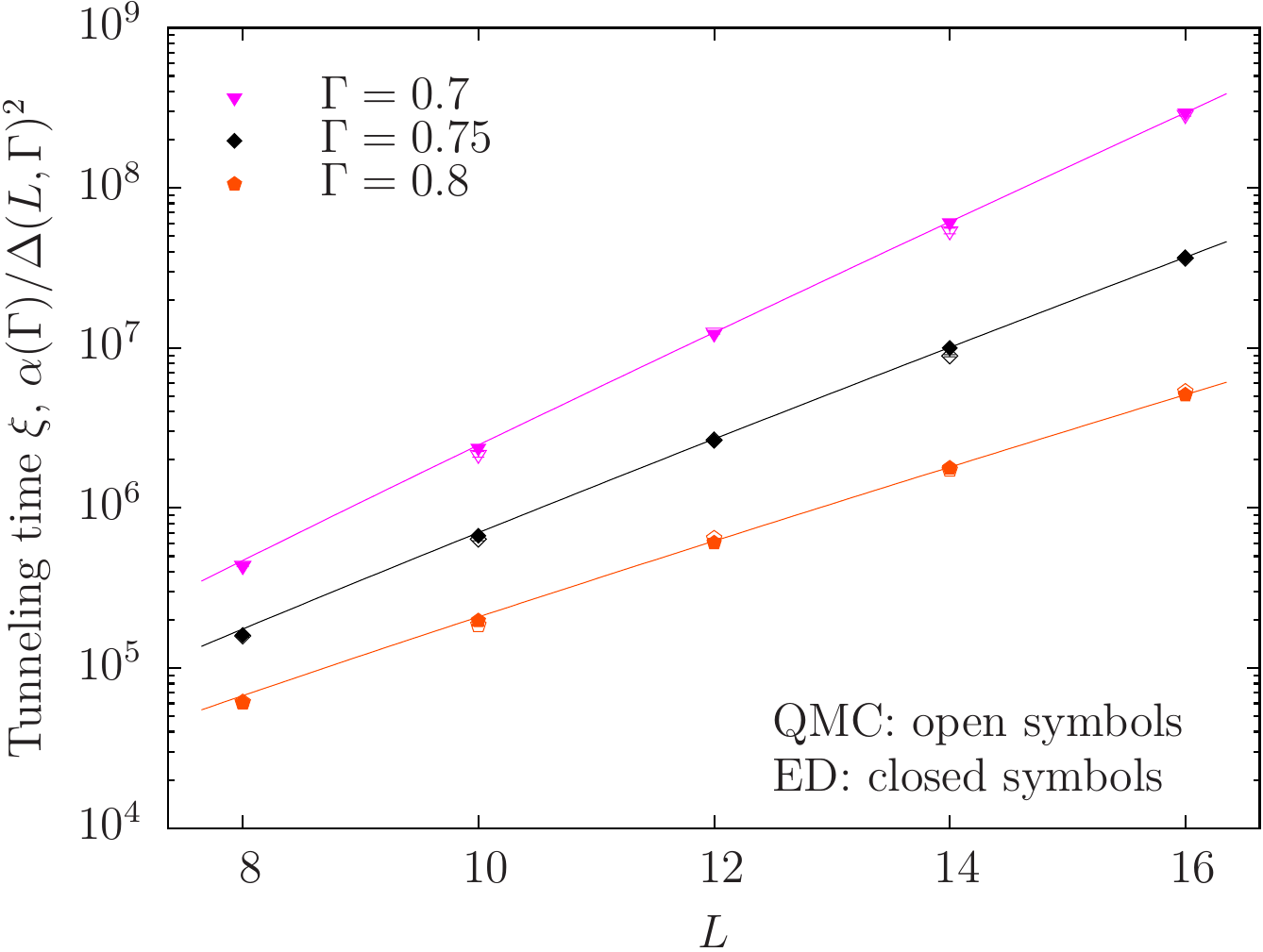}
  \caption{ (color online). Average PIMC tunneling time $\xi$ as a function of system size $L$ (open symbols) at $\beta=16$, for various values of the transverse field $\Gamma$. Exponential fits of the times for $12 \le L \le 16 $ are shown as solid lines. 
  To compare to physical QA we also show $a(\gamma)/\Delta(\Gamma,L)^2$, obtained by exact diagonalization (ED) with solid symbols. Rescaling the latter by $L$-independent constants $\alpha(\Gamma)$ we find identical scaling with system size $L$. 
 }
 \label{f:pbc6}
 \end{figure}

\begin{figure}[htp]
 \includegraphics[width=1\columnwidth]{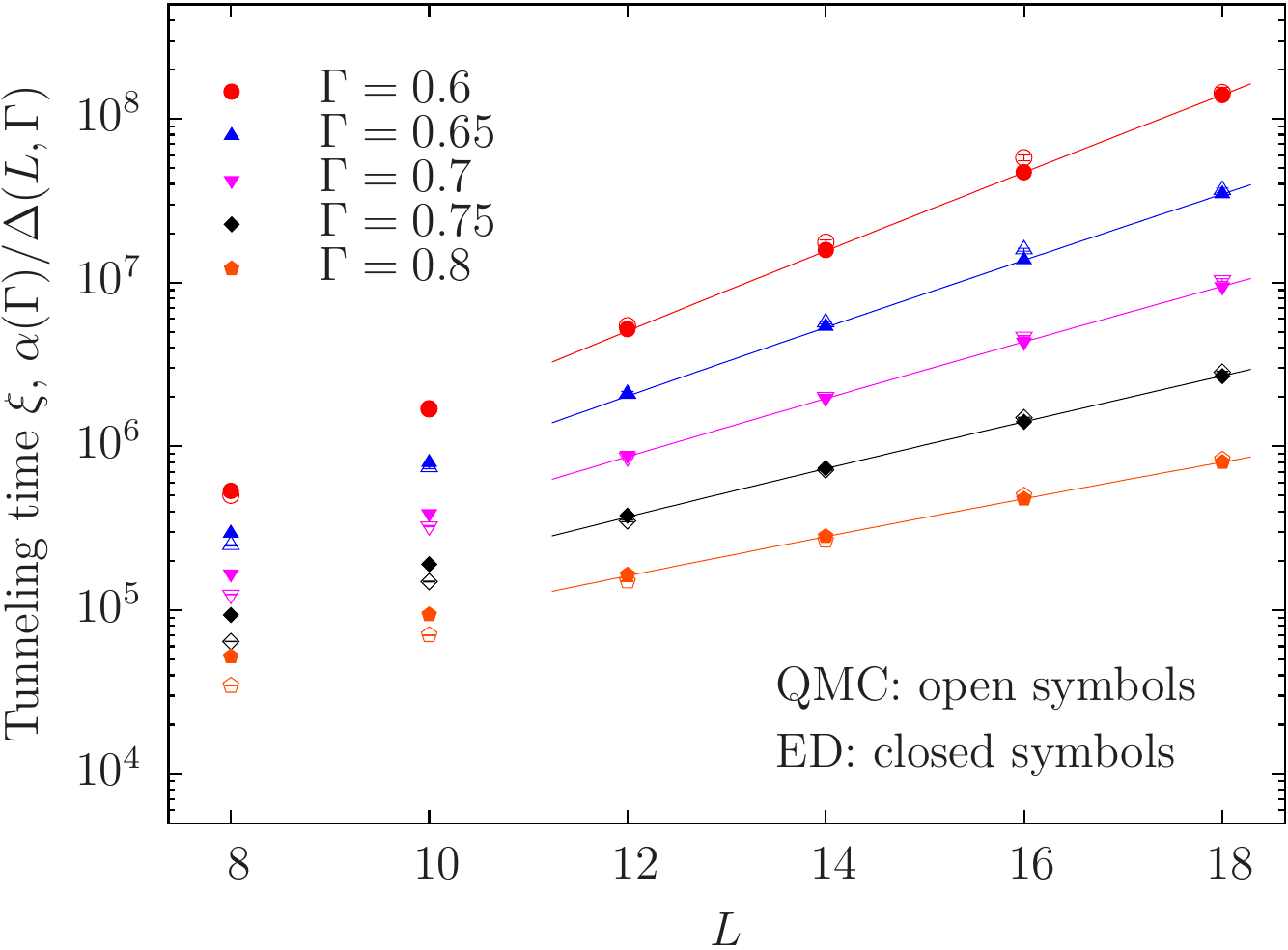}
 \caption{ (color online). Average PIGS tunneling  time $\xi$ as a function of  system size $L$ (open symbols) at $\beta=24$, for various values of the transverse field $\Gamma$. Exponential fits of the times for $12 \le L \le 18$ are shown as solid lines. Points proportional to $1/\Delta (L)$, obtained with exact diagonalization, are shown with filled symbols. }
 \label{f:obc4}
 \end{figure}

\begin{figure}[htp]
 \includegraphics[width=1\columnwidth]{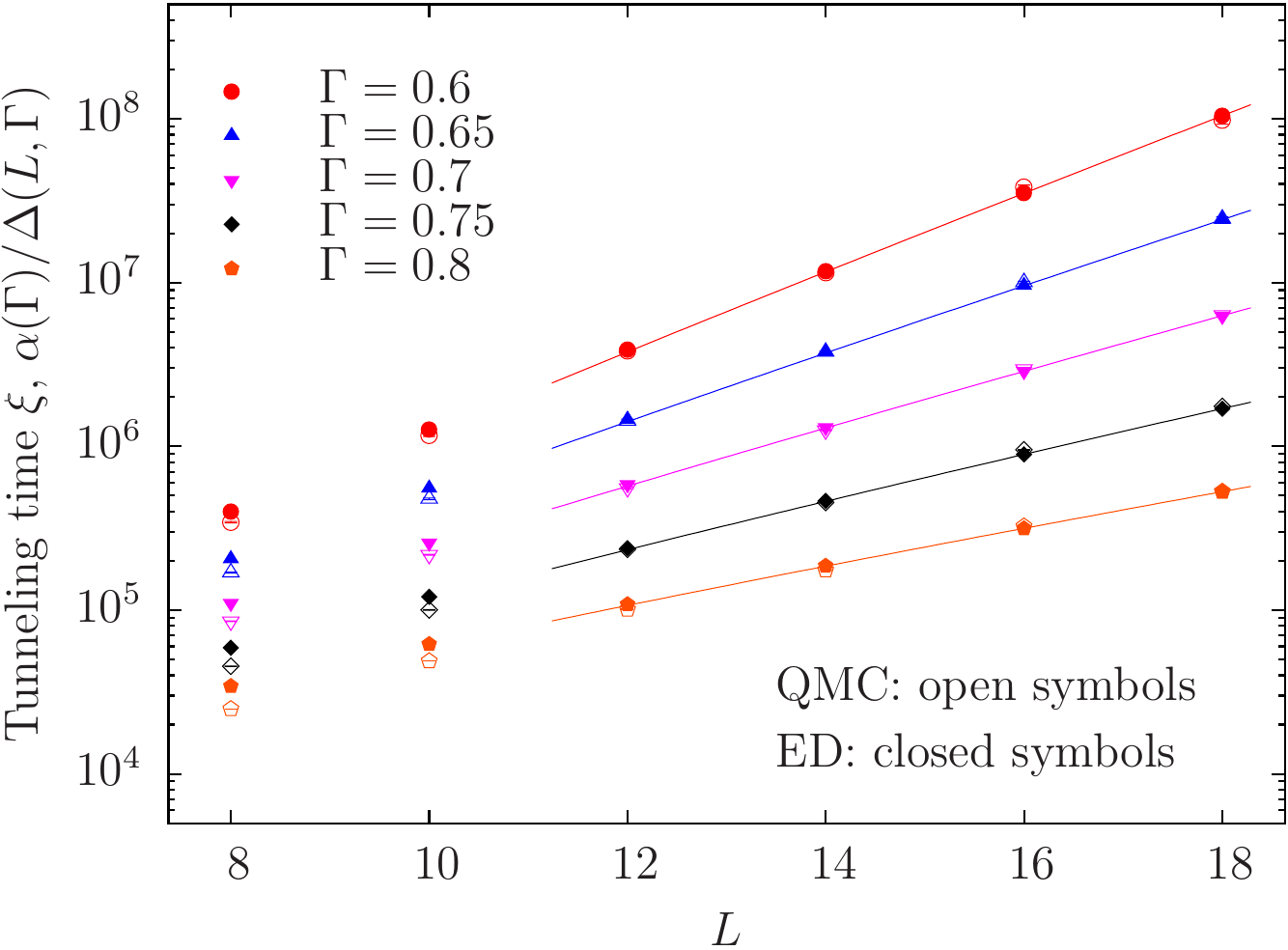}
  \caption{ (color online). Average PIGS tunneling  time $\xi$ as a function of  system size $L$ (open symbols) at $\beta=20$, for various values of the transverse field $\Gamma$. Exponential fits of the times for $12 \le L \le 18$ are shown as solid lines. Points proportional to $1/\Delta (L)$, obtained with exact diagonalization, are shown with filled symbols. }
 \label{f:obc5}
 \end{figure}

\begin{figure}[htp]
 \includegraphics[width=1\columnwidth]{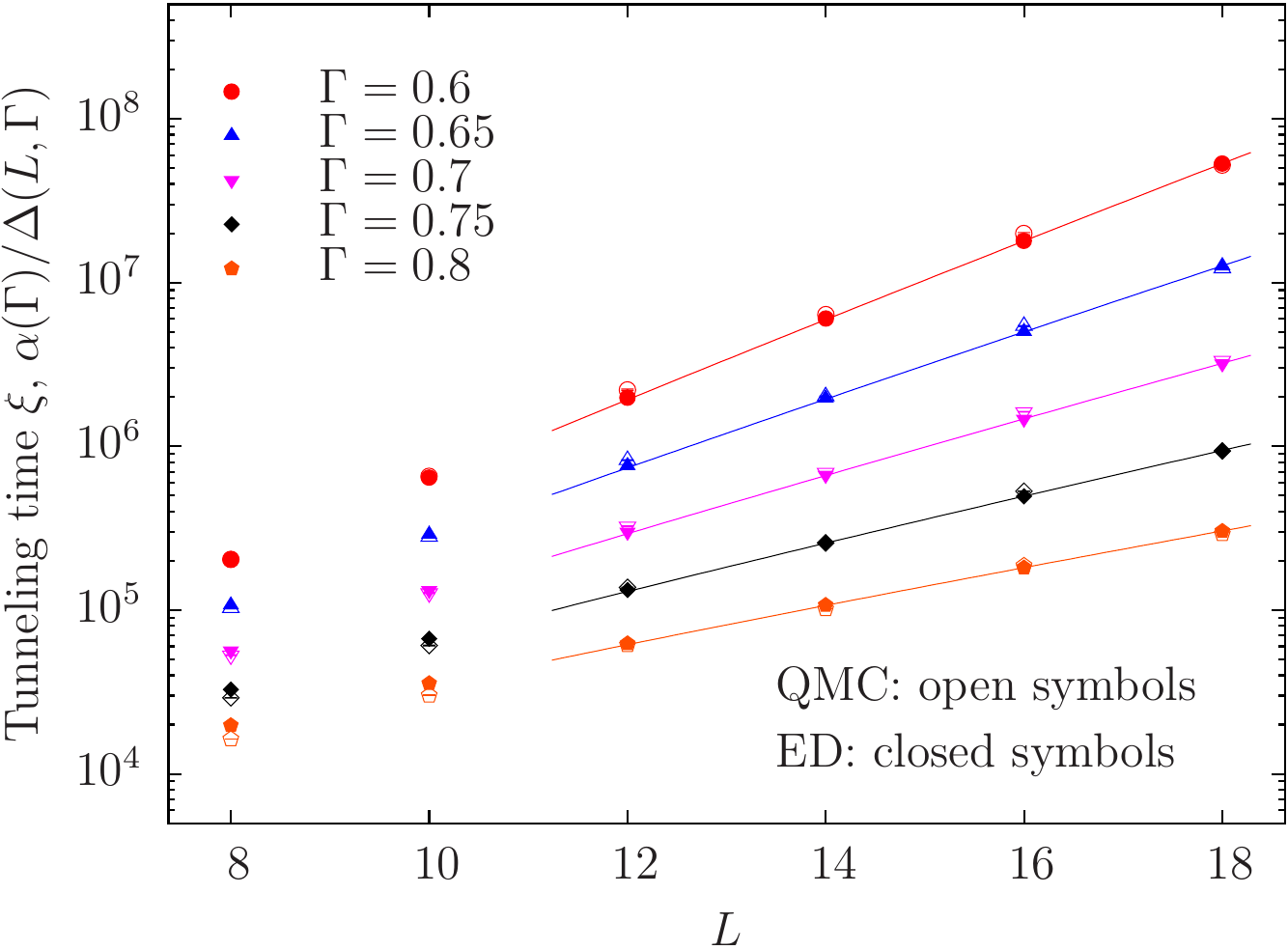}
  \caption{ (color online). Average PIGS tunneling  time $\xi$ as a function of  system size $L$ (open symbols) at $\beta=16$, for various values of the transverse field $\Gamma$. Exponential fits of the times for $12 \le L \le 18$ are shown as solid lines. Points proportional to $1/\Delta (L)$, obtained with exact diagonalization, are shown with filled symbols. }
 \label{f:obc6}
 \end{figure}

In Figs.~\ref{f:pbc4}, \ref{f:pbc5}, \ref{f:pbc6} we compare QMC simulations with periodic boundary conditions (PIMC), performed with different inverse temperatures $\beta = 24, 20,16$ and different transverse field values $\Gamma$, against the inverse squared gap $\Delta(\Gamma,L)^2$.
In Figs.~\ref{f:obc4}, \ref{f:obc5}, \ref{f:obc6} we compare instead QMC simulations with open boundary conditions (PIGS) against the inverse  gap $\Delta(\Gamma,L)$.
The complete exponential fit parameters are given in Table~\ref{ttable}.
The scaling of PIMC (PIGS) simulations compares well to $1/\Delta^2$ ($1/\Delta$) within error bars.

\begin{table}[ht]
\begin{center} 

\begin{tabular}{|c||c|c|c|c|} 
\hline  $\Gamma$  & $\beta=16$ &  $\beta=20$ & $\beta=24$ &ED   \\ 
\hline \hline & \multicolumn{3}{c|} {QMC-PIMC} &  $1/\Delta^2$  \\ 
\hline 0.8 & 0.52(2) &  0.553(17)  & 0.541(4) & 0.535(4) \\  
\hline 0.75 & 0.64(3) & 0.67(4)  & 0.6697(14) & 0.662(4) \\ 
\hline 0.7 & 0.78(2) & 0.80(3) & 0.81(5) & 0.799(4) \\  
\hline \hline &  \multicolumn{3}{c|} {QMC-PIGS} &   $1/\Delta$ \\ 
\hline 0.8 & 0.270(12) & 0.281(11) & 0.289(11) &  0.268(2)\\  
\hline 0.75  & 0.329(11) & 0.341(11)  & 0.350(9) & 0.331(2)\\ 
\hline 0.7  & 0.397(9) & 0.410(9) & 0.417(9) & 0.400(2)\\  
\hline 0.65  & 0.462(11)   & 0.477(8)  & 0.482(18)  & 0.473(2) \\
\hline 0.6 & 0.540(10) & 0.556(19) & 0.56(2) & 0.553(2)\\  
\hline \end{tabular} 
\end{center}
\caption{Exponents $b$ of the function $f(L) = a ~ \exp{(bL)}$ used to fit $\xi(L)$, $1/\Delta^2(L)$ and $1/\Delta(L)$, i.e. the average tunneling time, observed in PIMC and PIGS  simulations (see text), and the inverse gap obtained with exact diagonalization (ED). The error bar, for ED data are due to the scaling not perfectly following an exponential behavior. QMC error bars are dominated by statistical errors.}
\label{ttable}
\end{table}

    \begin{figure}[htp]
 \includegraphics[width=1\columnwidth]{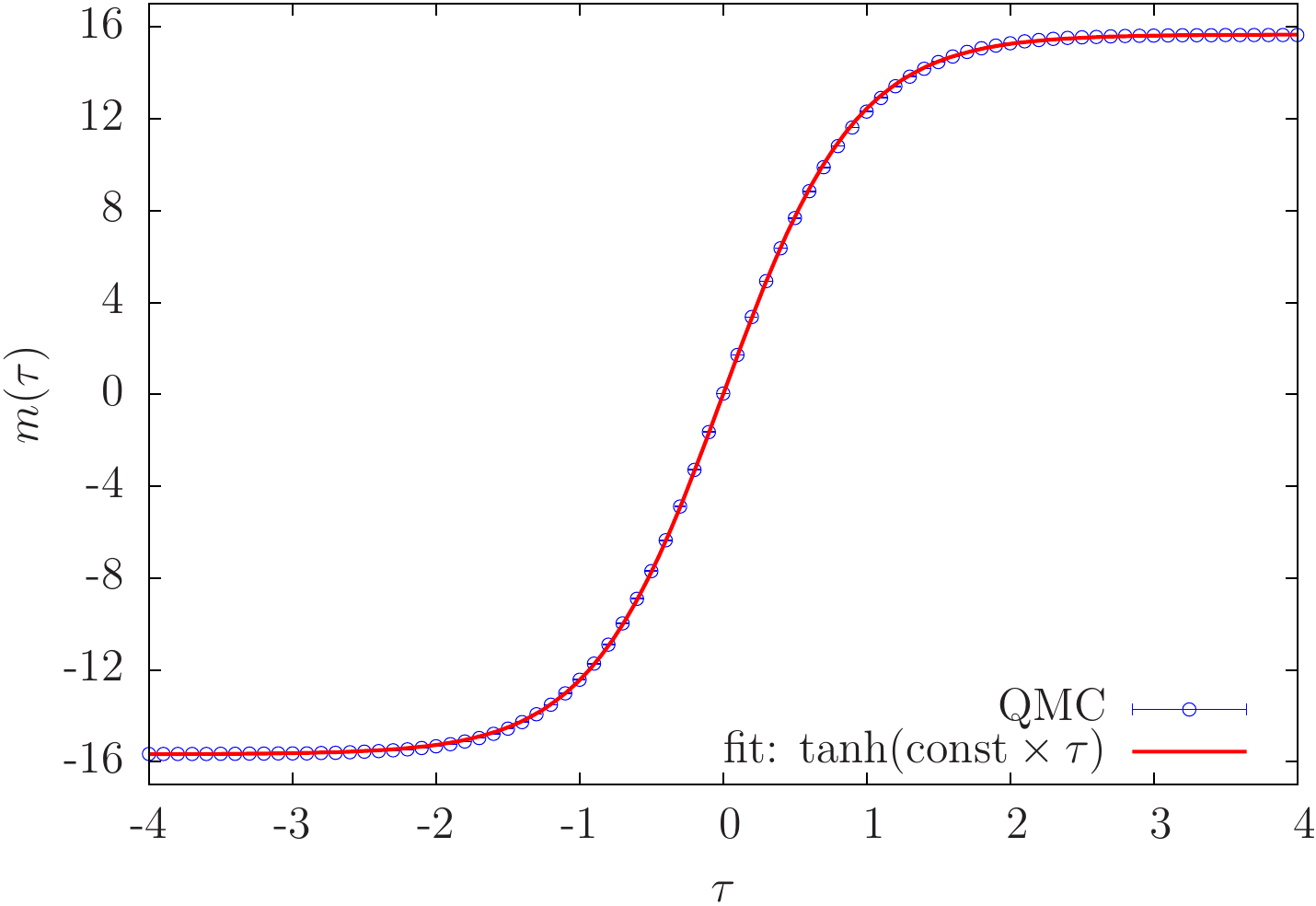} 
  \caption{ Instanton shape (QMC and analytical) for a ferromagnetic chain of size $L=16$ at $\Gamma=0.4$. We see that the instanton shape in QMC, given by $m(\tau)$ follows the analytical curve closely. }
 \label{f:shape}
 \end{figure}

\clearpage


\bibliography{bibQA}